\documentclass{emulateapj}
\submitted{Accepted for publication in ApJ on September 15, 2015}
\shorttitle{The origin of prolate rotation in dSph galaxies}

\shortauthors{I. Ebrov\'{a} and E. L. {\L}okas}

\begin{document}

\title{The origin of prolate rotation in dwarf spheroidal galaxies\\ formed by mergers of disky dwarfs}

\author{Ivana Ebrov\'{a}\altaffilmark{1} and Ewa L. {\L}okas\altaffilmark{2}}

\altaffiltext{1}{Astronomical Institute, The Czech Academy of Sciences, Bo\v{c}n\'{i} II 1401/1a,
CZ-141 00 Prague, Czech Republic}
\altaffiltext{2}{Nicolaus Copernicus Astronomical Center, Bartycka 18, PL-00-716 Warsaw, Poland}

%%%%%%%%%%%%%%%%%%%%%%%%%%%%%%%%%%%%%%%%%%%%%%%%%%%%%%%%%%%%%%%%%%%%%%%%%%%%%%%%
\begin{abstract}
Motivated by the discovery of prolate rotation of stars in Andromeda II, a dwarf spheroidal companion of M31,
we study
its origin via mergers of disky dwarf galaxies. We simulate merger events between two identical dwarfs changing the
initial inclination of their disks with respect to the orbit and the amount of orbital angular momentum. On radial
orbits the amount of prolate rotation in the merger remnants correlates strongly with the inclination of the disks and
is well understood as due to the conservation of the angular momentum component of the disks along the merger axis. For
non-radial orbits prolate rotation may still be produced if the orbital angular momentum is initially not much larger
than the intrinsic angular momentum of the disks. The orbital structure of the remnants with significant rotation is
dominated by box orbits in the center and long-axis tubes in the outer parts. The frequency analysis of stellar orbits
in the plane perpendicular to the major axis reveals the presence of two families roughly corresponding to inner and
outer long-axis tubes. The fraction of inner tubes is largest in the remnant forming from disks oriented most
vertically initially and is responsible for the boxy shape of the galaxy. We conclude that prolate rotation results
from mergers with a variety of initial conditions and no fine tuning is necessary to reproduce this feature. We compare
the properties of our merger remnants to those of dwarfs resulting from the tidal stirring scenario and the data for
Andromeda II.
\end{abstract}


%%%%%%%%%%%%%%%%%%%%%%%%%%%%%%%%%%%%%%%%%%%%%%%%%%%%%%%%%%%%%%%%%%%%%%%%%%%%%%%%
%%%%%%%%%%%%%%%%%%%%%%%%%%%%%%%%%%%%%%%%%%%%%%%%%%%%%%%%%%%%%%%%%%%%%%%%%%%%%%%%
\section{Introduction}

The most widely recognized theory of structure formation in the Universe is based on the assumption of the existence of
cold dark matter, significantly dominating the baryonic one. In such a Universe, galaxies form by hierarchical
merging of smaller systems. Large galaxies show plenty of signs of past or ongoing mergers but it seems harder to
collect observational evidence on the mass scale of dwarf galaxies. Nevertheless, some serious effort has already
been made in order to reveal signs of galactic mergers down to $\sim10^8$\,M$_{\sun}$ \citep{geh05, md12, crn14, pau15}.

In parallel, attempts have been made to understand dwarf mergers from the point of view of numerical simulations. In
this case high resolution is needed in order to adequately model a low-mass galactic merger in a cosmological context
because of the differences in scales. Although limited in resolution, studies in the context of the Local Group
demonstrated that a significant fraction of present-day dwarfs underwent a significant merger with another dwarf
galaxy in the past \citep{kli10, dea14}. These mergers most likely occur early on, at the outskirts of
the Local Group and between bound pairs or within groups. \citet{kaz11} demonstrated that mergers between initially
disky dwarf galaxies may produce pressure-supported systems with parameters very similar to those of dwarf spheroidal
(dSph) galaxies found in the Local Group. These systems seem particularly suitable for testing the merger paradigm on
small scales as their proximity offers a possibility for high-resolution measurements.

A few candidates for merger remnants have been already identified among the dSph galaxies of the Local Group.
The first was the Fornax dwarf for which shells of stars in the vicinity of the main body were observed \citep{col05}.
Other cases may include the most distant dSphs such as Cetus and Tucana that could not have interacted with the
Milky Way and M31 very strongly in the past and be tidally transformed so the merger scenario seems particularly
plausible \citep{kaz11}, though hypothetical as no direct observational evidence exists in this case.

%--------------------------------------
\begin{table*}
\begin{center}
\caption{Initial conditions for the simulations. }
\begin{tabular}{crrrrrrc}
\hline
\hline
Merger    &  $V_{\rm X} \ \ $ & $V_{\rm Y} \ \ $ & $V_{\rm Z} \ \ $
& $L_{\rm X}$/$L_{\rm tot}$ & $L_{\rm Y}$/$L_{\rm tot}$ & $L_{\rm Z}$/$L_{\rm tot}$
& Line type \\
          &   [km\,s$^{-1}$] \ & [km\,s$^{-1}$] \ & [km\,s$^{-1}$] \
&             &             &               &  \\
\hline
I90R  & $-8.0 $&  $ 0.0 $&  0.0 &  0.71 & 0.0 & $  0.71 $ & red solid    \\     % Merger4i90
      & $ 8.0 $&  $ 0.0 $&  0.0 &  0.71 & 0.0 & $ -0.71 $ & red dashed   \\

I60R  & $-8.0 $&  $ 0.0 $&  0.0 &  0.87 & 0.0 & $  0.50 $ & cyan solid  \\     % Merger5i60
      & $ 8.0 $&  $ 0.0 $&  0.0 &  0.87 & 0.0 & $ -0.50 $ & cyan dashed\\

I120R & $-8.0 $&  $ 0.0 $&  0.0 &  0.50 & 0.0 & $  0.87 $ & magenta solid \\         % Merger18i120
      & $ 8.0 $&  $ 0.0 $&  0.0 &  0.50 & 0.0 & $ -0.87 $ & magenta dashed\\

I90VY1 & $-8.0 $&  $  1.0$ &  0.0 &  0.71 & 0.0 &$  0.71 $ & green solid \\         % Merger14i90
       & $ 8.0 $&  $ -1.0$ &  0.0 &  0.71 & 0.0 &$ -0.71 $ & green dashed\\

I90VY2 & $-8.0 $&  $  2.0$ &  0.0 &  0.71 & 0.0 &$  0.71 $ & blue solid \\         % Merger16i90
       & $ 8.0 $&  $ -2.0$ &  0.0 &  0.71 & 0.0 &$ -0.71 $ & blue dashed\\
\hline
\label{tab:init}
\end{tabular}
\end{center}
\tablecomments{Odd rows list the initial conditions for Dwarf\,1 and even rows for Dwarf\,2 of a given merger.
The first column identifies the merger simulation (see text for details). The following columns list the
initial components of orbital velocities and angular momenta (in units of the total angular momentum of each dwarf).
The last column indicates the type and color of the line used for a given dwarf and
simulation in the figures throughout the paper.}
\end{table*}
%--------------------------------------

Recently, a new merger candidate appeared in the form of the Andromeda\,II (And\,II) dSph galaxy, one of the brightest
satellites of M31. The object has a number of peculiar features.
\citet{ho12} detected strong rotation signal along the 2D minor axis which means that the object
probably has a streaming motion around the longest axis in 3D, i.e. prolate rotation. Note that
And II is the only dSph with a clear prolate
rotation reported so far. \citet{mcc07} identified a few distinct stellar populations in And\,II
differing in age, metallicity and density profiles. However, \citet{ho12} found no significant difference
in the kinematics of the metal rich and metal poor stars.
The presence of the metallicity gradient was recently confirmed by \citet{ho15} and \citet{var14}.
\citet{var14} also found
an intermediate age population which makes And\,II an exception among their sample of six fainter dSphs that possess
only populations of stars older than 10\,Gyr. Moreover, \citet{am14} reported the detection of regions with
lower velocity dispersion in the kinematic data of \citet{ho12} which can be interpreted as a presence of a stellar
stream in And\,II.

\citet{lo14andii} proposed a merger scenario for the origin of And\,II. In this model, the merger occurs between two
dwarf galaxies initially composed of a disk and a dark matter halo. The dwarfs had equal masses and only differed
by the disk scale lengths. They were placed on a radial orbit towards each other and their disks were inclined
by $\pm45$\,deg to the collision axis so that the angular momentum vectors of the disks and the
velocities of the dwarfs were initially in the same plane. The model adequately reproduces the observed properties
of And II, including the prolate rotation and shape. The presence of the different stellar populations can be interpreted
as originating from different dwarfs: the two populations then have different density distributions (as a result of
different initial disk scale lengths) but very similar kinematics.

Another scenario known to lead to the formation of dSph galaxies is the one based on tidal interaction of the initially
disky dwarf satellites with bigger hosts, like the Milky Way or M31 \citep[e.g.][]{lo14bar, lo15}.
However, as argued by \citet{lo14andii}, tidal interaction is able to remove
very efficiently the initial rotation of the disk
(around the shortest axis) as it transforms to a bar and then a spheroid, but cannot induce any significant prolate
rotation (around the longest axis).

The initial conditions used in the simulation described in \citet{lo14andii} may seem rather special making the merger
highly improbable. In this paper we aim to investigate how much these
conditions may be relaxed and still reproduce the prolate rotation in the final product of the merger. We study the
effect of different initial inclination angles of the dwarf disks and the effect of non-radial orbits of the dwarfs.
In section 2 we present the simulations used in this study. The orbital evolution of the dwarfs is described
in section 3 while their shapes and kinematics in section 4. Section 5 is devoted to a more detailed study of
the orbital structure of the remnants. In section 6 we compare the predictions of the scenario based on mergers with
an alternative way to form dSph galaxies that relies on their interaction with a bigger host. In section 7 we compare
our predictions to the real data for And II and remark on the possible future, more complete model that will describe
all the available data. The summary and concluding remarks follow in section 8.

%%%%%%%%%%%%%%%%%%%%%%%%%%%%%%%%%%%%%%%%%%%%%%%%%%%%%%%%%%%%%%%%%%%%%%%%%%%%%%%%
%%%%%%%%%%%%%%%%%%%%%%%%%%%%%%%%%%%%%%%%%%%%%%%%%%%%%%%%%%%%%%%%%%%%%%%%%%%%%%%%

\section{Simulations}

We performed five collisionless simulations of mergers between two identical dwarf galaxies.
The simulations differed in the initial inclinations of the disks in the dwarfs
and their relative velocities. The numerical realization of the dwarf galaxy is the same as used
previously in simulations of tidal stirring by \citet{lo14bar} and \citet{lo15}. The model
was generated following the method described in \citet{wid05} and \citet{wid08} which allows to
create self-consistent, equilibrium $N$-body models of galaxies with an exponential disk and an
NFW halo \citep{nfw97}.

Our dwarf galaxy had a dark matter component in the form of
the NFW halo of total mass $M_{\mathrm{h}}=10^9$\,M$_{\sun}$ and concentration
parameter $c=20$. The exponential disk had the mass of $M_{\mathrm{d}}=2\times10^7$\,M$_{\sun}$, the scale-length
$R_{\mathrm{d}}=0.41$\,kpc and thickness $z_{\mathrm{d}}/ R_{\mathrm{d}}=0.2$.
Each component of the dwarf contained $10^6$ particles (which makes the total of $4\times10^6$ particles per simulation).
The mergers were followed using the $N$-body simulation code GADGET-2 \citep{sp05} with the adopted softening scale
lengths of $0.02$\,kpc and
$0.06$\,kpc for the disk and halo particles, respectively. Each merger was evolved for 10\,Gyr with positions and
velocities of the particles stored every 0.05\,Gyr.

We chose the coordinate system of the simulation box so that its origin was at the center of gravity of the two dwarfs.
Dwarf\,1 was always placed at coordinates $(X,Y,Z)=(-25,0,0)$\,kpc and Dwarf\,2 at $(25,0,0)$\,kpc. For radial mergers
$X$ is the original collision axis,
for the other ones $XY$ is the original collision plane. In order to investigate the effects of different disk
inclinations and
merger orbits, we keep the same model of mass distribution for both dwarfs in all simulations. We also keep the same
initial distances, 50\,kpc, and the same relative velocity in the direction of the $X$ axis, 16\,km\,s$^{-1}$. At the
beginning of each simulation, the angular momentum vectors of the dwarf disks lie in the $XZ$ plane. Of course, due to
the finite precision of the computation, the dwarfs can slightly deviate from a given axis or plane during the
simulations.

The components of the initial angular momenta of the dwarf disks and the relative initial
velocities of the dwarfs for all
five simulations are listed in Table~\ref{tab:init}. The labels assigned to the simulations
(the first column of the Table) refer to the starting configuration: the
number after `I' stands for the inclination angle (in degrees) between the angular momentum vectors of the dwarfs,
`R' means radial merger, and
for non-radial mergers `VY' indicates the velocity (in km\,s$^{-1}$) along the $Y$ axis.

We use capital letters to denote relative velocities or distances of the two dwarfs in the simulation as well as the
original coordinate system we defined for the simulation ($X,Y,Z$). Small letters ($x,y,z$) refer to the coordinate
system associated with the principal axes of the components representing the luminous matter of either individual dwarfs
(Dwarf\,1 or Dwarf\,2) or of the whole galaxy resulting from the merger (i.e. disk components of both dwarfs combined).
In the latter system the major axis corresponds to the $x$ axis, the intermediate one to $y$ and the minor one to $z$.
Similarly, other quantities
related to this system are marked with small letters, e.g. $v_x$ is the rotation around the major axis of the
respective dwarf or galaxy, and such quantities always refer to just the luminous matter.

For comparisons of the results we divide the simulations into two groups: (1) radial merger
simulations with different initial
inclinations of the disks (I90R, I60R and I120R) and (2) simulations with the same initial inclination but with
different initial relative velocities (I90R, I90VY1 and I90VY2). The simulation I90R is included in both groups
for reference and is treated as default. This simulation is closest to the one discussed by \citet{lo14andii}
and shown to reproduce well the properties of And II, except for the fact the we now use identical dwarf progenitors.

%%%%%%%%%%%%%%%%%%%%%%%%%%%%%%%%%%%%%%%%%%%%%%%%%%%%%%%%%%%%%%%%%%%%%%%%%%%%%%%%
%%%%%%%%%%%%%%%%%%%%%%%%%%%%%%%%%%%%%%%%%%%%%%%%%%%%%%%%%%%%%%%%%%%%%%%%%%%%%%%%

%--------------------------------------
\begin{figure}
\centering{}
\resizebox{\hsize}{!}{\includegraphics[angle=-90]{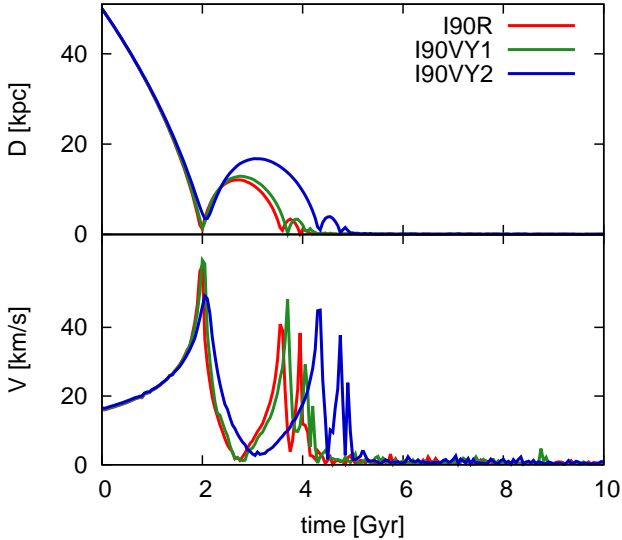}}
\caption{Relative distances (top panel) and velocities (bottom panel) of the dwarfs for simulations with different
initial relative velocities.}
\label{fig:disvel}
\end{figure}
%--------------------------------------

%--------------------------------------
\begin{figure}
\centering{}
\vspace{-4.5mm}
\resizebox{\hsize}{!}{\includegraphics[angle=-90]{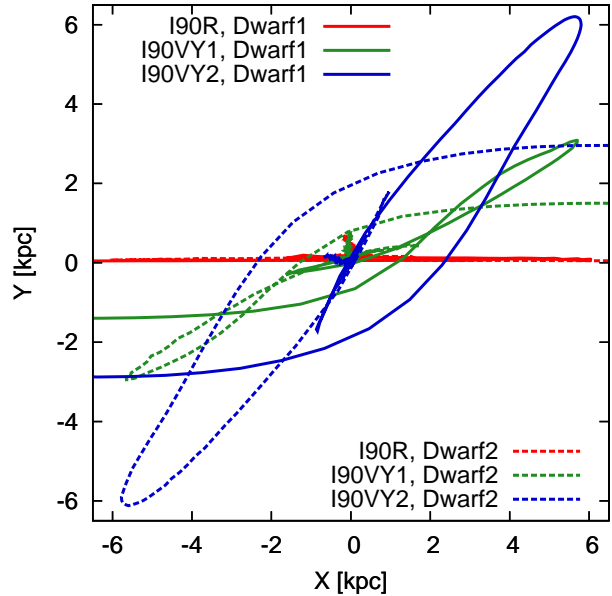}}
\caption{Trajectories of dwarfs in simulations with different initial relative velocities.}
\label{fig:orbits}
\end{figure}
%--------------------------------------

\section{Orbits of merging dwarfs}

Fig.~\ref{fig:disvel} shows the evolution of relative distances and velocities of the
dwarfs for simulations with different
initial relative velocities. For these simulations, the trajectories of the dwarfs in the collision plane
(the $XY$ plane)
are displayed in Fig.~\ref{fig:orbits}. For our radial mergers (along the $X$ axis), the orbits are almost the same
in all three cases.

In the case of radial mergers the initial conditions and the subsequent evolution are very symmetrical.
As a result, the major axis of the merger remnant is aligned with the merger axis (the $X$ axis of the simulation box).
By introducing some velocity in
the $Y$ direction (for simulations I90VY1 and I90VY2) we introduce asymmetry in our simulations because the merger
becomes prograde for Dwarf\,1 and
retrograde for Dwarf\,2. This causes differences in the properties of the two dwarfs at the end of the simulation for
non-radial mergers. All dwarfs initially have the velocity of 8\,km\,s$^{-1}$ towards the common center of mass.
Simulations I90VY1 and I90VY2 have additional 1\,km\,s$^{-1}$ and  2\,km\,s$^{-1}$, respectively, in the perpendicular
($Y$) direction. The second approach of the dwarfs is on a close-to-radial orbit in both cases but for I90VY2
it takes about 0.7\,Gyr longer to reach it.

For I90VY1 the merging times and results are quite similar to simulation I90R, although
there is less rotation in the final product. The major axis of the merger remnant is in the $XY$ plane
but inclined by about 22\,deg with respect to the $X$ axis.
For I90VY2 the major axis of the remnant also stays in the $XY$ plane, now inclined by 67\,deg to the $X$ axis,
but otherwise the simulation outcome is very different. There is just a little rotation in the final product
and there are significant differences
in the shape and rotation between Dwarf\,1 and Dwarf\,2 at a given time step. For this reason, in the following figures
showing the properties of the dwarfs, we plot just the results for Dwarf\,1 except for simulation I90VY2 for which
we show results for Dwarf\,2 as well (the blue dashed curves).

%--------------------------------------
\begin{figure*}
\centering{}
\resizebox{\hsize}{!}{\includegraphics[angle=-90]{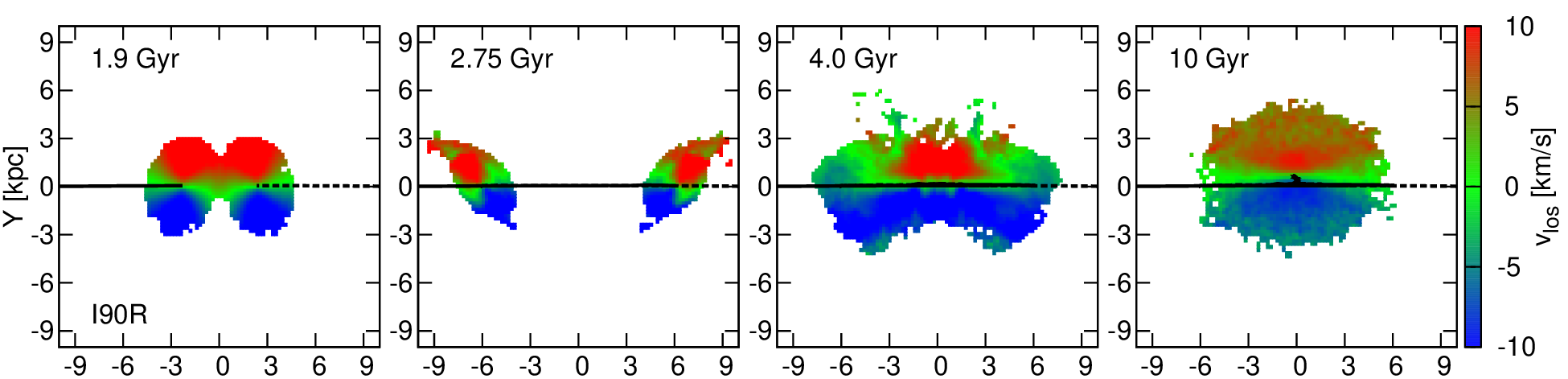}}\\
\resizebox{\hsize}{!}{\includegraphics[angle=-90]{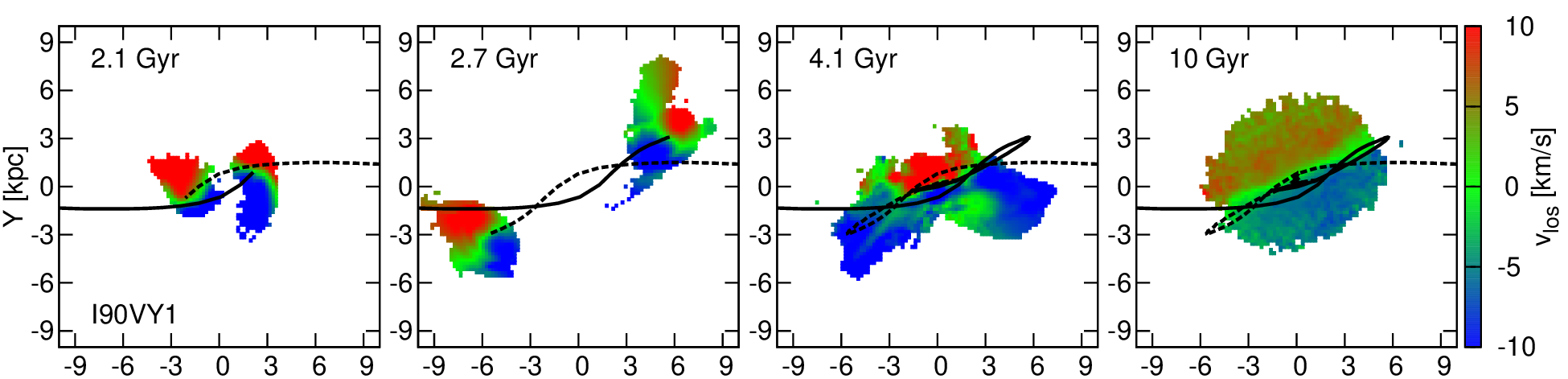}}\\
\resizebox{\hsize}{!}{\includegraphics[angle=-90]{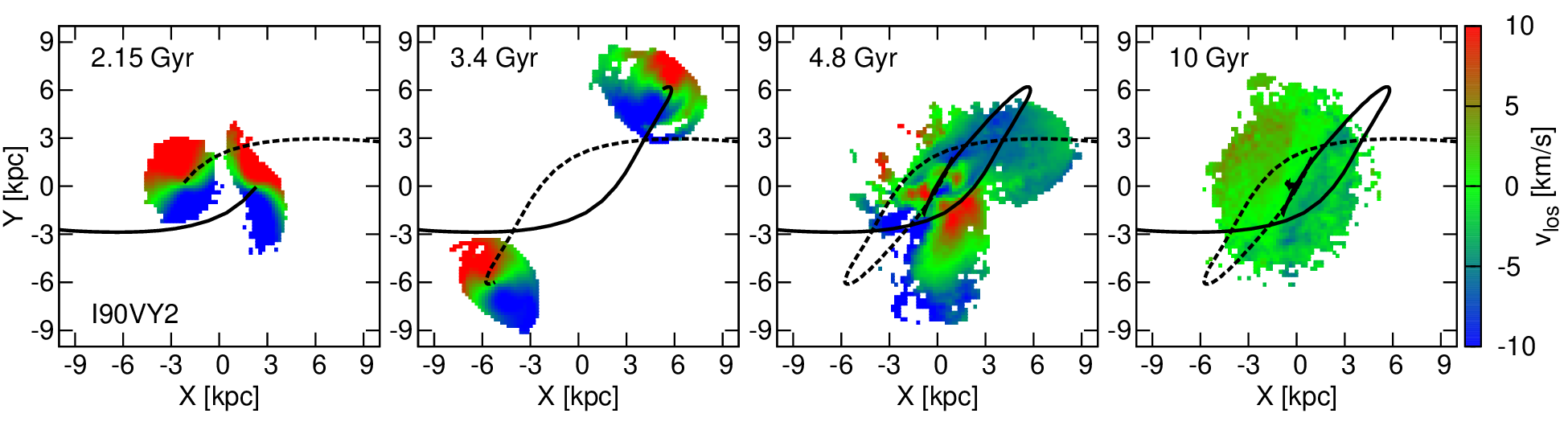}}
\caption{Snapshots from the simulations with different initial relative velocities.
The three rows of panels show dwarfs from simulation I90R, I90VY1 and I90VY2 respectively,
in projection onto the orbital plane $XY$. Panels in different columns correspond to different stages of evolution,
as marked in the upper left corner of each plot.
Black lines indicate the orbits followed by the dwarfs up to a given time with the solid line for Dwarf\,1 and
the dashed one for Dwarf\,2, respectively. The colors code the
mean line-of-sight velocity. Each panel shows the area of $20\times20$\,kpc.}
\label{fig:vmaps}
\end{figure*}
%--------------------------------------

Fig.~\ref{fig:vmaps} shows several snapshots from each of the simulations with different initial relative velocities.
The dwarfs are shown in projection onto the orbital plane $XY$ and the colors code the
mean line-of-sight velocity. The first panel of each
row shows the situation around the first pericenter passage. We note that Dwarf\,1 of I90VY2 (on the right side of the
panel) is notably more deformed than Dwarf\,2. In the second-column panels dwarfs are located near their first
apocenters and the third-column snapshots roughly correspond to the time when the centers of the dwarfs have
just merged, but the merger remnant is still far from equilibrium. The last-column panels are the final snapshots
from the simulations, after 10 Gyr of evolution.

%%%%%%%%%%%%%%%%%%%%%%%%%%%%%%%%%%%%%%%%%%%%%%%%%%%%%%%%%%%%%%%%%%%%%%%%%%%%%%%%
%%%%%%%%%%%%%%%%%%%%%%%%%%%%%%%%%%%%%%%%%%%%%%%%%%%%%%%%%%%%%%%%%%%%%%%%%%%%%%%%

%--------------------------------------
\begin{figure}
\centering{}
\vspace{2mm}
\resizebox{\hsize}{!}{\includegraphics[angle=-90]{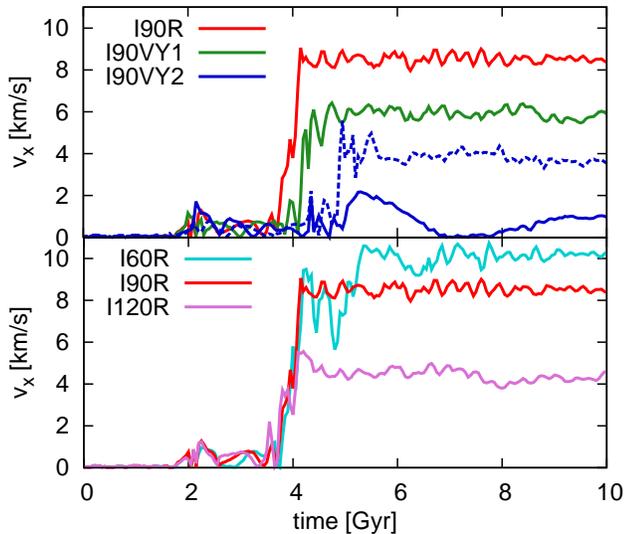}}
\caption{Evolution of the mean rotation velocity around the major axis for Dwarf\,1 in the simulations with
different initial relative velocities (top panel) and for the radial mergers with different inclination of the disks
(bottom panel). The blue dashed curve corresponds to Dwarf\,2 from simulation I90VY2.}
\label{fig:rot}
\end{figure}
%--------------------------------------

%--------------------------------------
\begin{figure}
\centering{}
\resizebox{\hsize}{!}{\includegraphics[angle=-90]{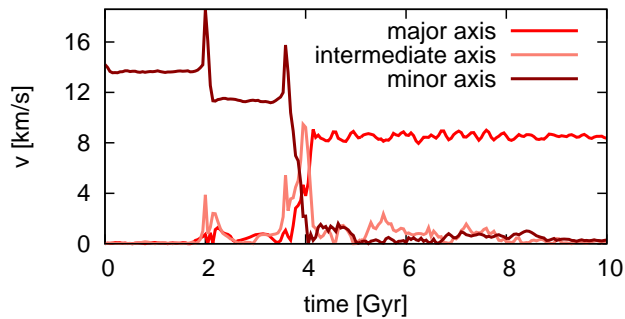}}
\caption{Evolution of the mean rotation velocities around the principal axes of Dwarf\,1 in simulation I90R.}
\label{fig:rot4i90}
\end{figure}
%--------------------------------------

\section{Kinematics and shape}
\subsection{Prolate rotation} \label{sub:rot}

In order to check how much prolate rotation is present in each merger remnant,
we found the principal axes of the stellar component of each dwarf separately and computed the
mean rotation velocity around the major axis in the sphere of radius 2\,kpc centered on the center of the
respective dwarf. We
plot the evolution of this rotation in time for Dwarf\,1 for all simulations in Fig.~\ref{fig:rot}.

Due to the symmetry of the initial conditions in radial mergers, the evolution of this rotation in Dwarf\,1 is the
same as for Dwarf\,2 in the same simulation. For I90VY1, the asymmetry is too small to lead to significant
differences between the two dwarfs. On the other hand, in the case of I90VY2 the evolution is
dramatically different for the two dwarfs, so we include the result for Dwarf\,2 as a dashed curve in
Fig.~\ref{fig:rot}.

For all simulations, there is essentially no rotation around the intermediate axis and the evolution of
the rotation around the minor axis is trivial: the disk preserves the rotation until the two dwarfs actually start to
merge and then the rotation vanishes and remains close to zero until the end of the simulation. We show one
example of the evolution of the mean rotation velocity around all principal axes for our default simulation
I90R in Fig.~\ref{fig:rot4i90}.

%--------------------------------------
\begin{figure}
\centering{}
\resizebox{\hsize}{!}{\includegraphics[angle=-90]{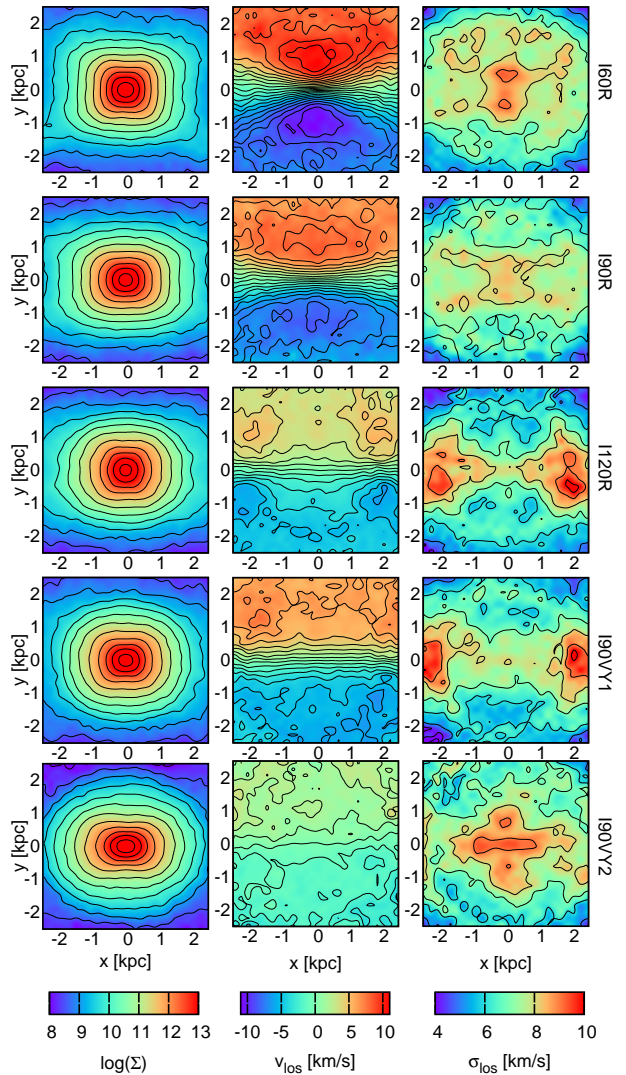}}
\caption{Maps of the surface density (first column), the mean line-of-sight velocity (middle column) and the
line-of-sight velocity dispersion (last column) for the stars in the
final outputs of the simulations as viewed along the minor
axis of the combined stellar component. Different rows correspond to different simulations, as marked on the right.
More vertical orientations of the initial disks and more radial merger orbits lead to stronger prolate rotation.
The surface density of the stars is in units of 645 M$_{\sun}$\,kpc$^{-2}$ with contours spaced by
0.5 in $\log \Sigma$. The contours in velocity and velocity dispersion are spaced by 1 km s$^{-1}$.}.
\label{fig:maps}
\end{figure}
%--------------------------------------

Fig.~\ref{fig:maps} shows the maps of line-of-sight properties of the final outputs for all five simulations
in projection onto the $xy$ plane, i.e. as seen along the $z$ axis of the merger remnant.
Here, $x$, $y$ and $z$ correspond to the major, intermediate and minor axis of
the combined stellar components of both dwarfs. The images confirm that the rotation signal (middle-column panels of
Fig.~\ref{fig:maps}) is around the longest ($x$) axis, i.e. the rotation is clearly prolate in all simulations,
even when the rotation level is very low as in I90VY2.

The amount of prolate rotation in the final product of a given simulation seems to be controlled in a
simple way. For radial mergers, decreasing the initial angle between the angular momenta of the
dwarfs and the merger axis (which corresponds to decreasing the angle between the angular momenta of the two dwarfs,
from 120 to 90 and 60 deg) we get more rotation in the final result. When the initial angular momenta of the disks are
more aligned with the merger axis the components of the momenta along the merger axis are larger. Since
these components are preserved during such symmetrical mergers the final products have more prolate rotation.

For non-radial mergers, the more orbital velocity we add in the direction perpendicular to the radial,
the less intrinsic rotation we get. This result can be interpreted by comparing the orbital and angular momenta
of the dwarfs. In the case of I90VY1 (I90VY2) the orbital angular momentum of each dwarf is about twice (four times)
larger than the intrinsic angular momentum of the stellar component of the dwarf.
We conclude that the prolate rotation in the remnant cannot be produced if the orbital angular momentum during the
merger is significantly larger than the intrinsic one.

The maps of line-of-sight velocity
dispersion in the right column of Fig.~\ref{fig:maps} show considerable variation among the different simulations
without any clear trend: some remnants have a maximum dispersion in the center while others have the maxima at
larger distances. It seems that the complicated structure of the dispersion maps may be used as an indicator of
the merger origin of the galaxy even after many Gyr of evolution.

%--------------------------------------
\begin{figure}
\centering{}
\vspace{2mm}
\resizebox{\hsize}{!}{\includegraphics[angle=-90]{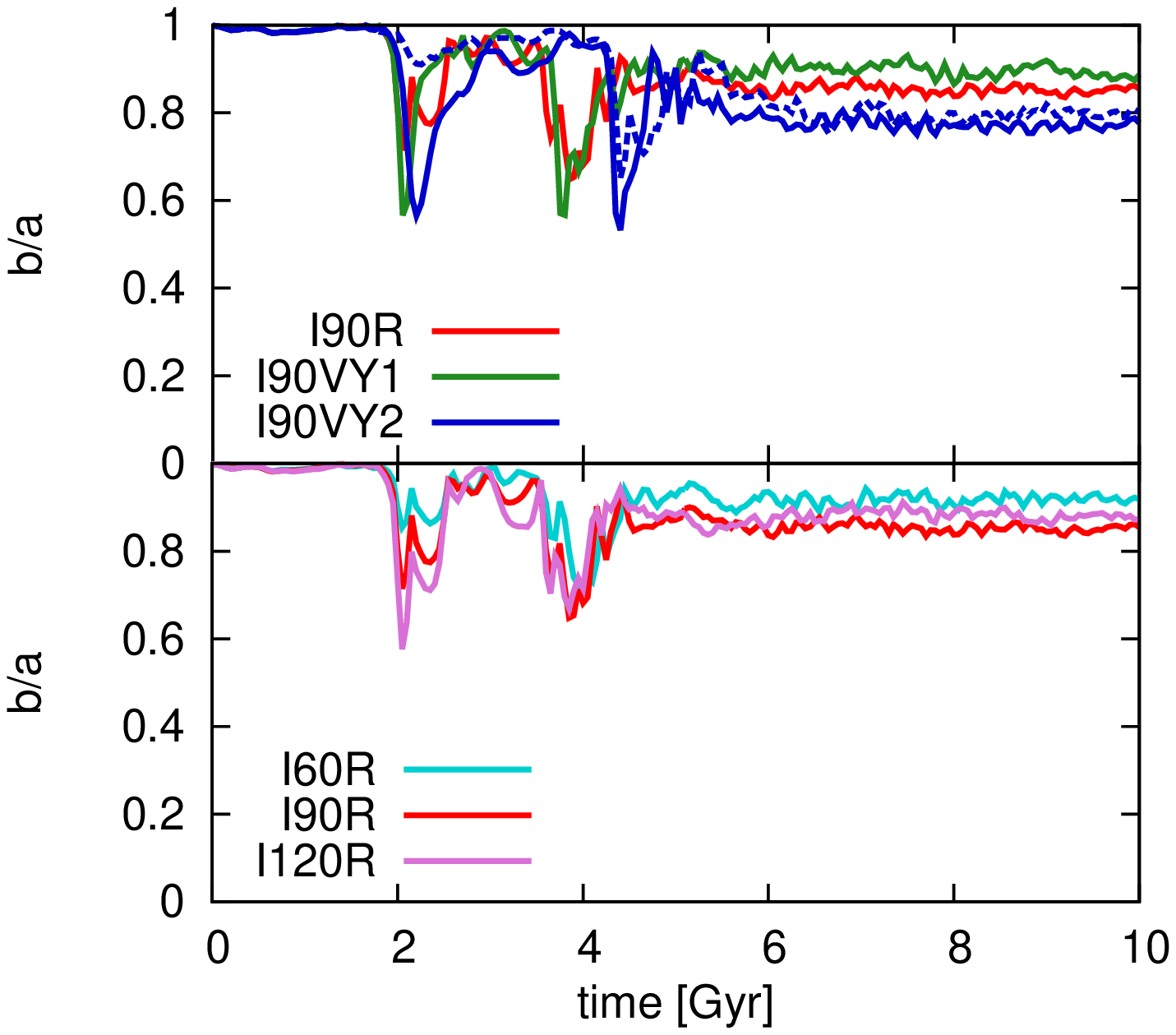}}\\
\resizebox{\hsize}{!}{\includegraphics[angle=-90]{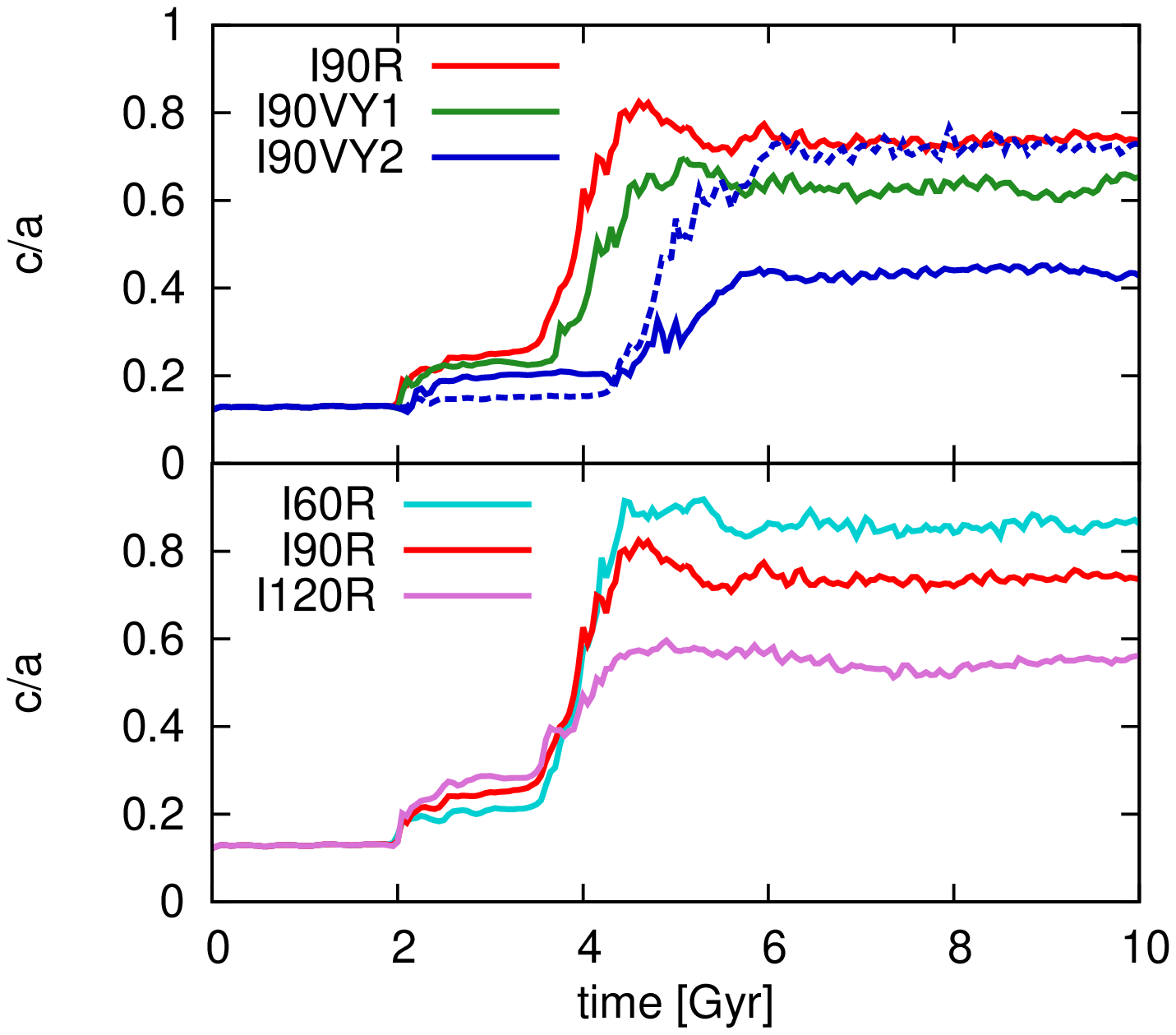}}\\
\resizebox{\hsize}{!}{\includegraphics[angle=-90]{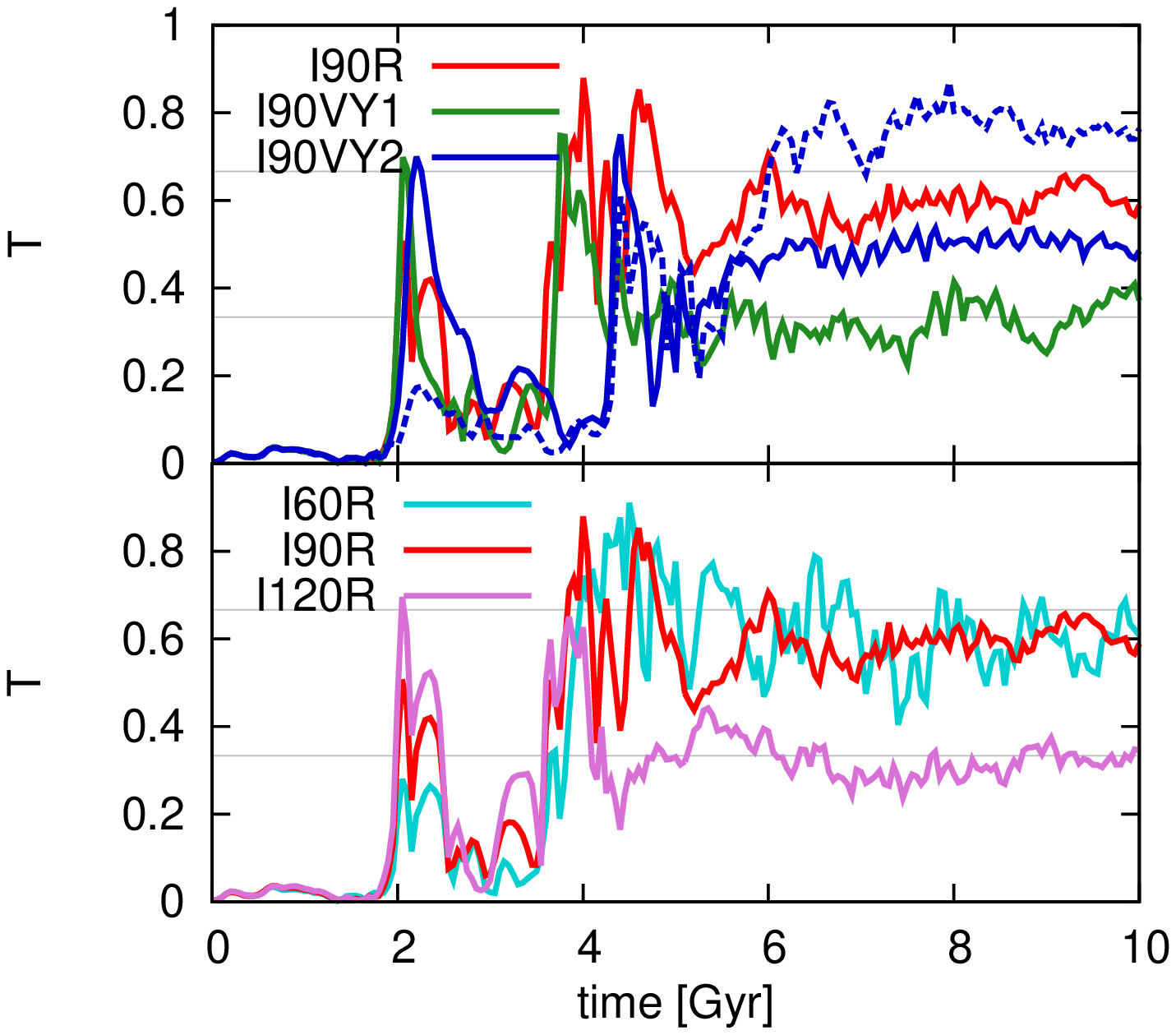}}
\caption{Evolution of the axis ratios and the triaxiality parameter for Dwarf\,1 in different simulations. The gray
horizontal lines in the lower two panels separate the regions of oblate, triaxial and prolate shape.
The blue dashed curve corresponds to Dwarf\,2 of simulation I90VY2.}
\label{fig:triax}
\end{figure}
%--------------------------------------
%%%%%%%%%%%%%%%%%%%%%%%%%%%%%%%%%%%%%%%%%%%%%%%%%%%%%%%%%%%%%%%%%%%%%%%%%%%%%%%%
 \subsection{Triaxiality}

In order to study the evolution of the shape of the merger remnants, we again analyzed each dwarf separately
selecting the stars inside the sphere of 2\,kpc centered on a
given dwarf. In Fig.~\ref{fig:triax} we plot the intermediate to major axis ratio, $b/a$, the minor to major axis ratio,
$c/a$, and the triaxiality parameter, $T$, defined as
$T=[1-(b/a)^2]/[1-(c/a)^2]$. When $T$ is lower than $1/3$ the galaxy is rather oblate, values greater than $2/3$
indicate a prolate shape and when $T$ falls between these values the galaxy has a triaxial shape.

%--------------------------------------
\begin{figure}
\centering{}
\vspace{6mm}
\resizebox{\hsize}{!}{\includegraphics[angle=-90]{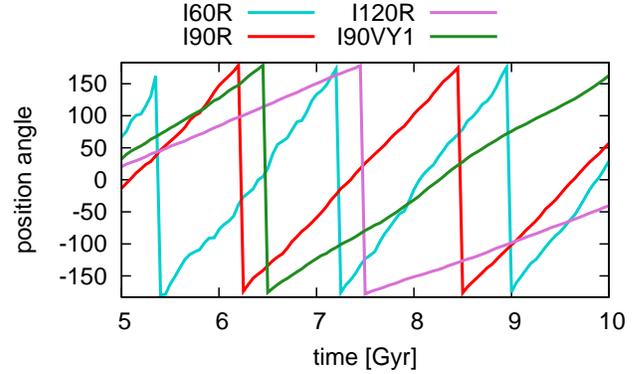}}
\caption{Position angles of the intermediate axes of the stellar components for the four simulations which exhibit
figure rotation. The angle is measured in the $YZ$ plane during the last 5\,Gyr of each simulation. }
\label{fig:PAs}
\end{figure}
%--------------------------------------

The most dramatic change of the shape of the dwarfs happens around the second passage (approximately at
$\sim$4\,Gyr). The ratio $b/a$, which is equal to unity at the beginning, slightly drops to values between about 0.8 and
0.9 for all simulations. The evolution of $c/a$ is more interesting. It grows from the original value of
0.12 up to the value ranging
from 0.43 for I90VY2 (Dwarf\,1) to 0.86 for I60R. For radial mergers, there is a simple rule for the behavior of $c/a$:
the final ratio decreases when the initial angle between the angular momenta of the dwarfs increases, i.e. the merger
remnants become thinner. For merger
I90VY2 $b/a$ remains similar for both dwarfs but $c/a$ is significantly different (0.43 for Dwarf\,1 and 0.73 for
Dwarf\,2). In the later stages of the simulations, all dwarfs can be classified as triaxial. Only I90VY1 crosses
the line of the oblate shape in several time steps and Dwarf\,2 of I90VY2 has more prolate values of triaxiality.

Referring again to Fig.~\ref{fig:maps} we note that the shapes of the merger remnants also differ in terms of their
boxiness. In this respect, the dependence seems to be the strongest for the radial mergers.
Clearly, the simulation with the dwarf disks oriented most vertically (I60R) with respect to the merger axis
the isodensity contours are
most boxy. In order to make this statement more quantitative we calculated the Fourier mode $A_4$ of the surface
distributions of the stars (such as those shown in the left column of Fig.~\ref{fig:maps}) for all simulations.
The amplitude of the $m$th Fourier mode is given as $A_m = (1/N) \left| \Sigma^{N}_{j=1} \exp (i m \phi_j) \right|$
where $\phi_j$ are the particle phases in the projection plane and $N$ is the total number of particles.
It turns out that over the last few Gyr of evolution $A_4\approx 0.06$ for I60R while it is at least a factor of
two lower for the other simulations. This shape must originate from the orbital structure of the merger remnant and
we discuss this issue further in the following sections.

%%%%%%%%%%%%%%%%%%%%%%%%%%%%%%%%%%%%%%%%%%%%%%%%%%%%%%%%%%%%%%%%%%%%%%%%%%%%%%%%
%%%%%%%%%%%%%%%%%%%%%%%%%%%%%%%%%%%%%%%%%%%%%%%%%%%%%%%%%%%%%%%%%%%%%%%%%%%%%%%%

%--------------------------------------
\begin{figure}
\raggedright{}
\vspace{4mm}
\resizebox{0.95 \hsize}{!}{\includegraphics[angle=-90]{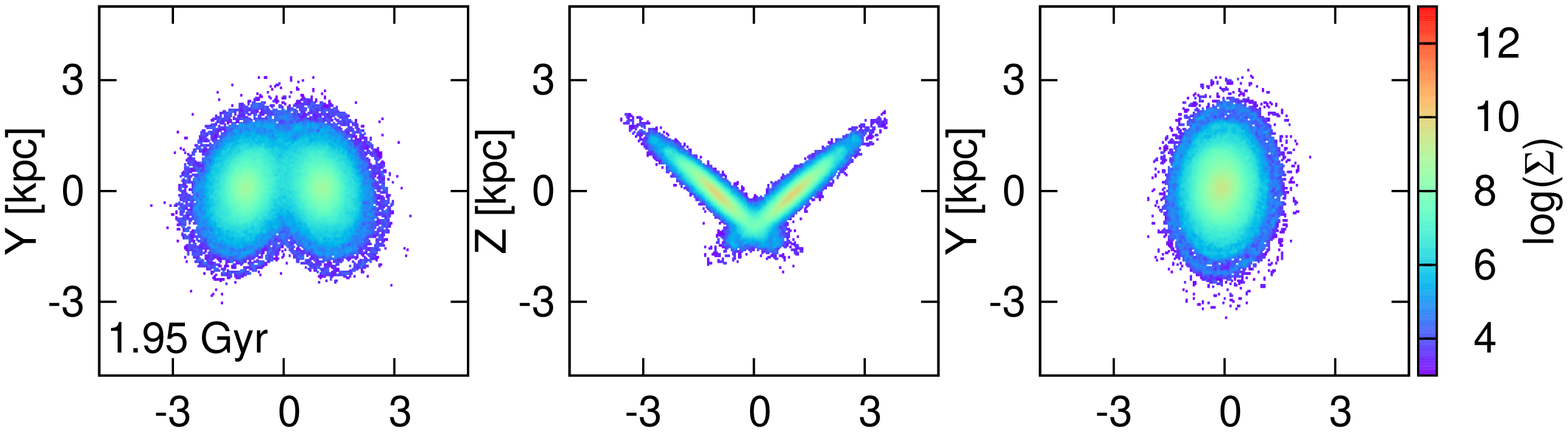}}\\
\resizebox{0.95 \hsize}{!}{\includegraphics[angle=-90]{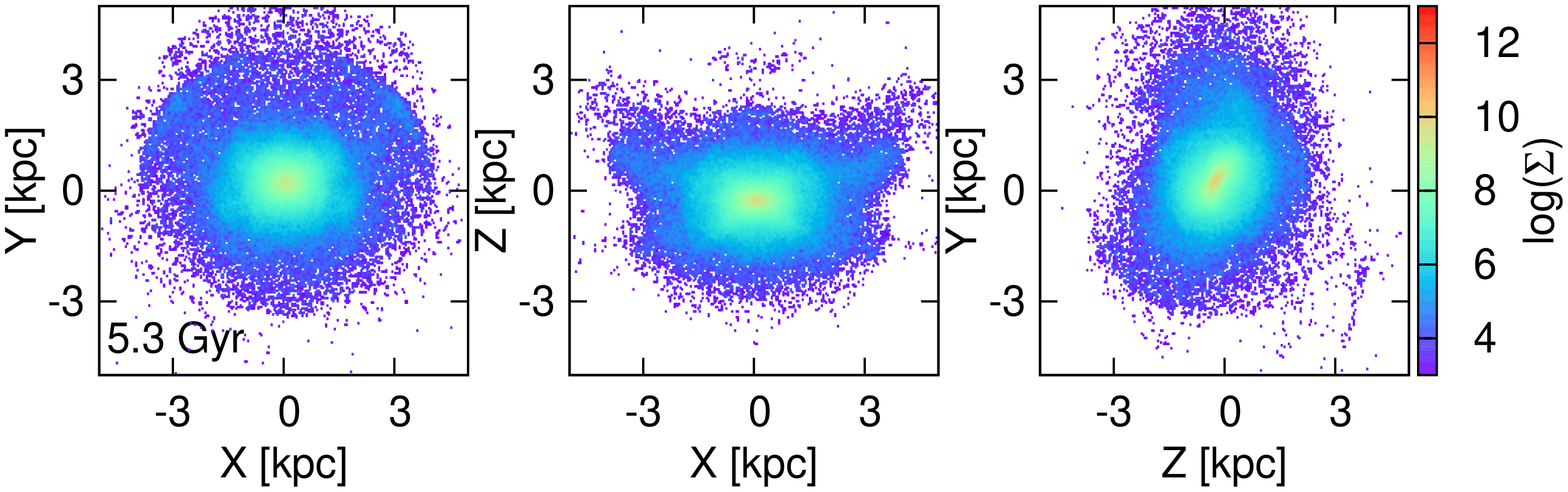}}
\caption{Surface density of stellar particles in simulation I90R at two time steps: at the first encounter (upper row)
and after the merger (lower row). Columns show projections along the $Z$, $Y$ and $X$ axis of the simulation box
(from the left to the right).}
\label{fig:2maps3d}
\end{figure}
%--------------------------------------

\section{Orbital structure}

\subsection{Figure rotation}

Although the ratios of principal axes do not vary significantly over the last 5\,Gyr for all simulations,
all but I90VY2 show clear
figure rotation around the major $x$ axis. The orientation of the major axis remains stable for the whole period of
5\,Gyr and it
coincides with the collision ($X$) axis for radial mergers and is inclined by 22\,deg from the collision axis for
I90VY1 but remains in the $XY$ plane. The evolution of the position angles of the intermediate axes for these four
simulations are plotted in Fig.~\ref{fig:PAs}. For the purpose of these measurements we computed
the principal axes for the combined stellar
components from both dwarfs in each simulation.

The periods of rotation of the intermediate (and short) axes are 1.7, 2.2, 6.0, and 3.7\,Gyr for I60R, I90R, I120R,
and I90VY1, respectively. The speed of the figure rotation is correlated with the rotation of stellar particles around
the long axis (see Fig.~\ref{fig:rot}) but the particles rotate at a higher speed (at least for the inner 2\,kpc).
Also the direction of the particle rotation is the same as the figure rotation.

The origin of the figure rotation can be traced to the tidal distortion of the disk just before the first encounter.
We illustrate this in Fig.~\ref{fig:2maps3d} which displays stellar density in simulation I90R in three projections
along the $Z$, $Y$ and $X$ axis of the simulation box. The upper row shows the projections just before the first
passage at 1.95\,Gyr and the bottom row illustrates the situation soon after the two disks merged, at 5.3\,Gyr.
The upper left panel of the Figure (view along the $Z$ axis)
shows the distortion of the disks due to the tidal force acting before the
merger. Since the stellar particles in both disks are moving upwards in this plot in the parts of the disks that are
already in touch, the tidal force distorted the disk so that the upper parts are closer than the lower ones. This
results in the tumbling (clockwise around the $X$ axis) of the elongated shape after the merger seen in the lower
right panel (view along the $X$ axis).

Simulation I90VY2 has the major axis inclined by 67\,deg with respect to the $X$ axis and lies in the $XY$ plane
also being stable for the last 5\,Gyr of evolution. The position angles of the other axes vary by only about 30\,deg
over the last 3 Gyr so this merger remnant does not show any significant figure rotation.

%%%%%%%%%%%%%%%%%%%%%%%%%%%%%%%%%%%%%%%%%%%%%%%%%%%%%%%%%%%%%%%%%%%%%%%%%%%%%%%%

%--------------------------------------
\begin{figure}
\centering{}
\vspace{-5mm}
\resizebox{\hsize}{!}{\includegraphics[angle=-90]{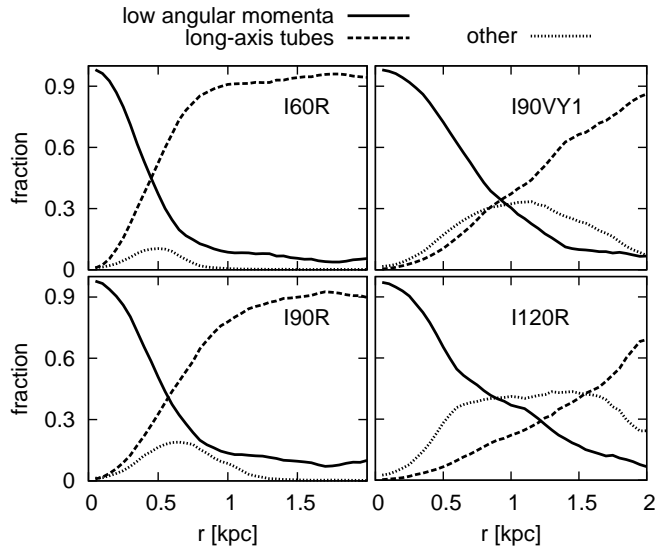}}
\caption{Fraction of a given orbit type as a function of radius for the stellar component in the final output of
simulations I60R, I90R, I120R and I90VY1. Different line types correspond to different orbit types as shown
in the legend.}
\label{fig:orbfra}
\end{figure}
%--------------------------------------

\subsection{Orbit classification}

In this subsection, we focus on simulations which exhibit significant prolate rotation, so we exclude I90VY2 from
further analysis.

At the beginning of the simulations stellar particles are situated in inclined disks on nearly circular orbits.
During the merger particle orbits are transformed to form an object with a shape that can be well approximated by
a triaxial ellipsoid. In static triaxial potentials most particle orbits belong to one of a few major families:
box orbits,
short-axis tube orbits, and inner and outer long-axis tube orbits \citep{dez85}. \citet{schw82} also constructed
a model of the triaxial ellipsoid with figure rotation around the short axis by properly populating different orbit
families. Classifying orbits in a non-static potential is more difficult.

We performed a simplified classification of particle orbits using the last 4\,Gyr of the simulations when the
galaxies are already merged. We defined one orientation of the coordinate system ($x,y',z'$) which coincides
with the actual principal axes of the galaxy only at one or few particular moments. For the rest of the time only
the major axis is along the $x$ coordinate.

We adopted modified criteria of orbit classification from \citet{schw93} as recently used by \citet{gaj15}.
Particles that keep a constant sign of the $x$-component of the angular momentum, $L_x$, are marked as
\textit{long-axis tubes}. Since the direction of the short axis is not stable, short-axis tubes are not
classified in our case.

We also find orbits with low average values of the $x$, $y'$, and $z'$-component of the angular momentum.
Specifically,
we require the sum of these average values, $L_{xy'z'}$, to be less than 0.1\,M$_{\sun}$\,kpc$^{2}$\,Myr$^{-1}$ (for
comparison, in all outputs almost all particles inside 2\,kpc have current total angular momentum smaller than
1\,M$_{\sun}$\,kpc$^{2}$\,Myr$^{-1}$). When the figure rotation is present the box orbits acquire some net angular
momentum \citep{schw82} thus we call this group just \textit{low angular momenta\/} orbits.
It is worth noting that $L_{y'}$
and $L_{z'}$ of most of the particles average out to zero and $L_{xy'z'}$ is typically dominated by the
average value of $L_x$.

The rest of the particles, which do not fall into one of the above categories, are marked as \textit{other}.
These particles have
$L_{xy'z'}>0.1$\,M$_{\sun}$\,kpc$^{2}$\,Myr$^{-1}$ and at the same time they do not keep a constant sign of $L_x$
for the duration of the last 4\,Gyr of the simulation.

Fig.~\ref{fig:orbfra}  shows fractions of the three types of orbits as a function of radius in the final outputs of the
simulations. The fraction of long-axis tubes correlates tightly with the amount of the mean rotation
velocity of
particles discussed in section~\ref{sub:rot}. Galaxies with higher rotation have faster increase of the fraction of
long-axis tubes with radius, mainly at the expense of the low angular momenta orbit family.

%--------------------------------------
\begin{figure}
\centering{}
\vspace{4mm}
\resizebox{\hsize}{!}{\includegraphics[angle=-90]{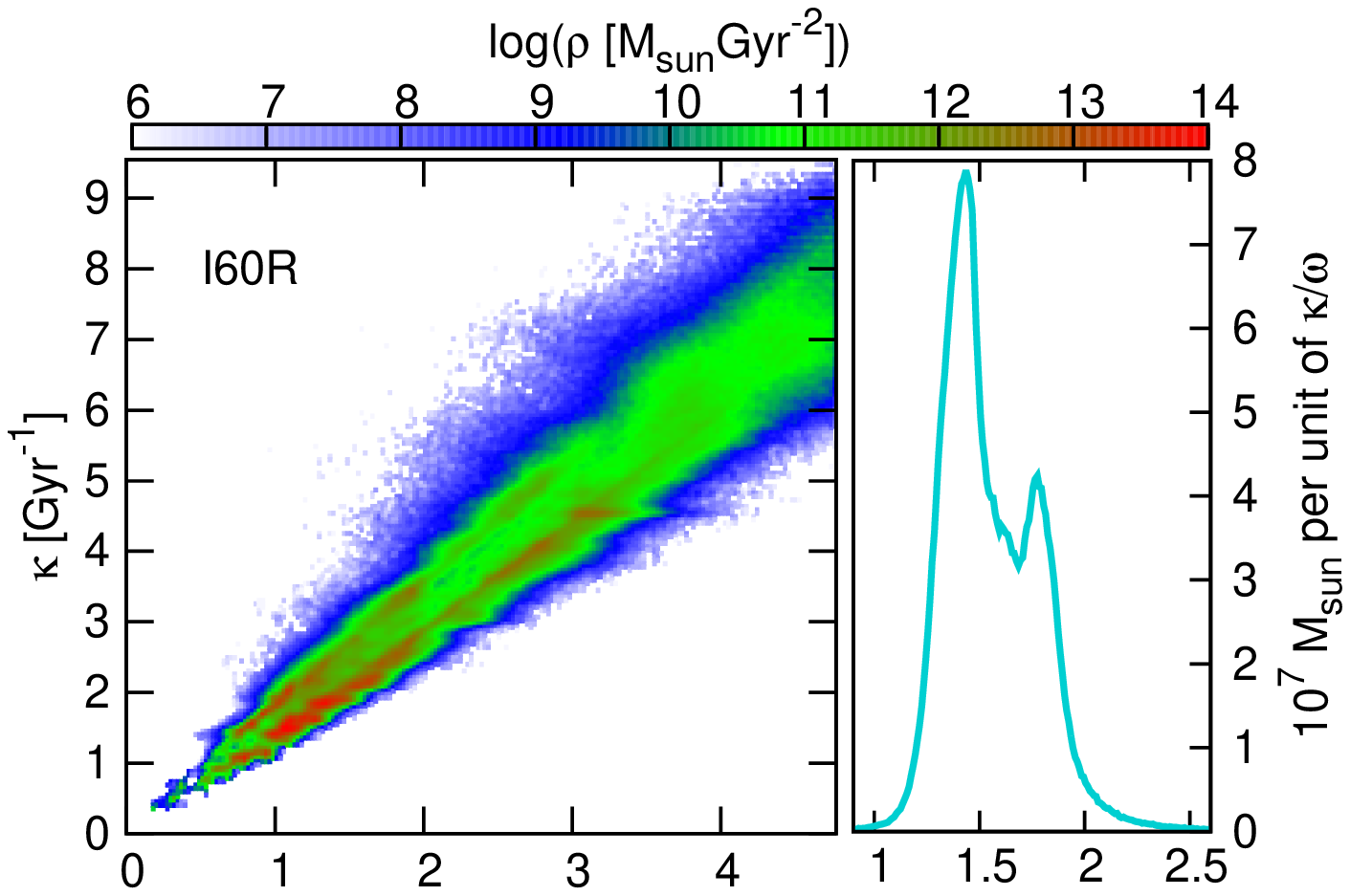}}\\
\resizebox{\hsize}{!}{\includegraphics[angle=-90]{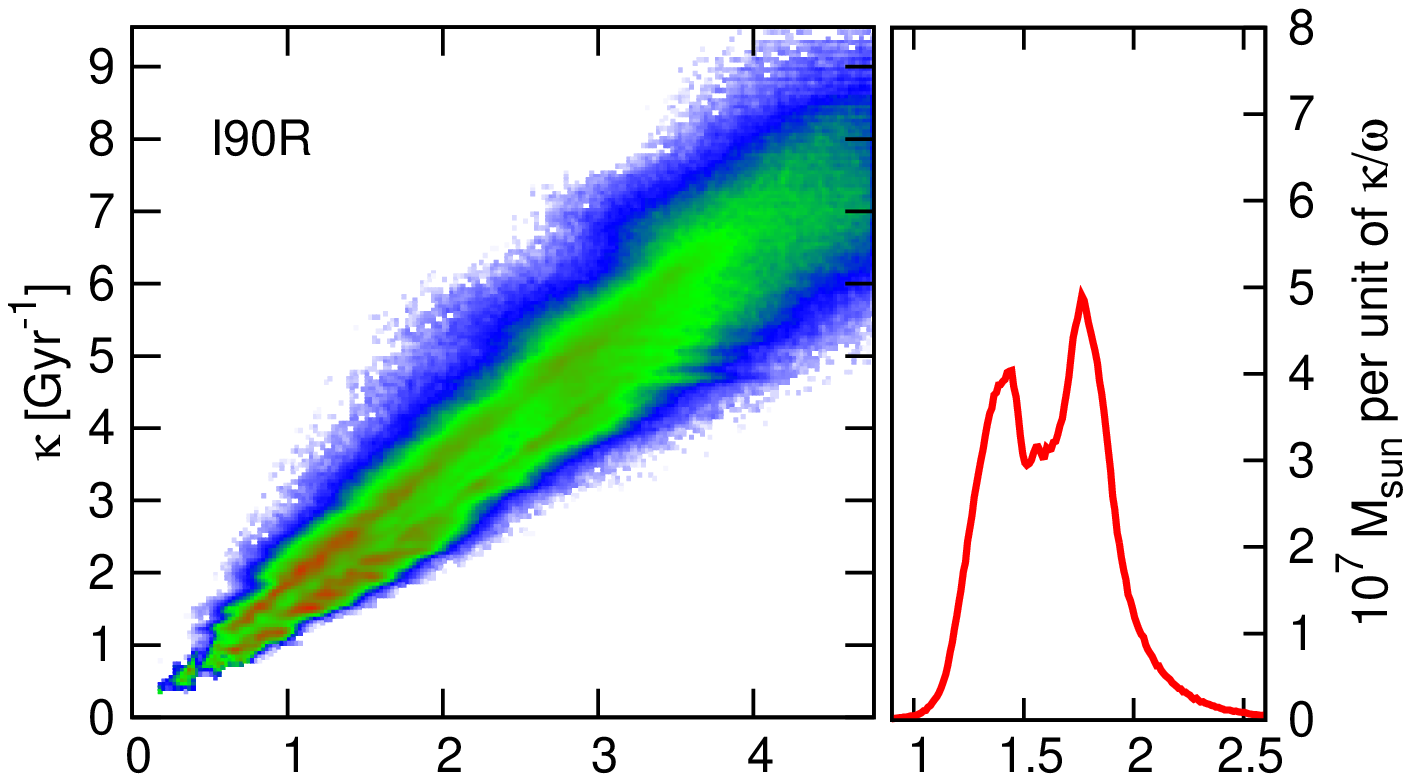}}\\
\resizebox{\hsize}{!}{\includegraphics[angle=-90]{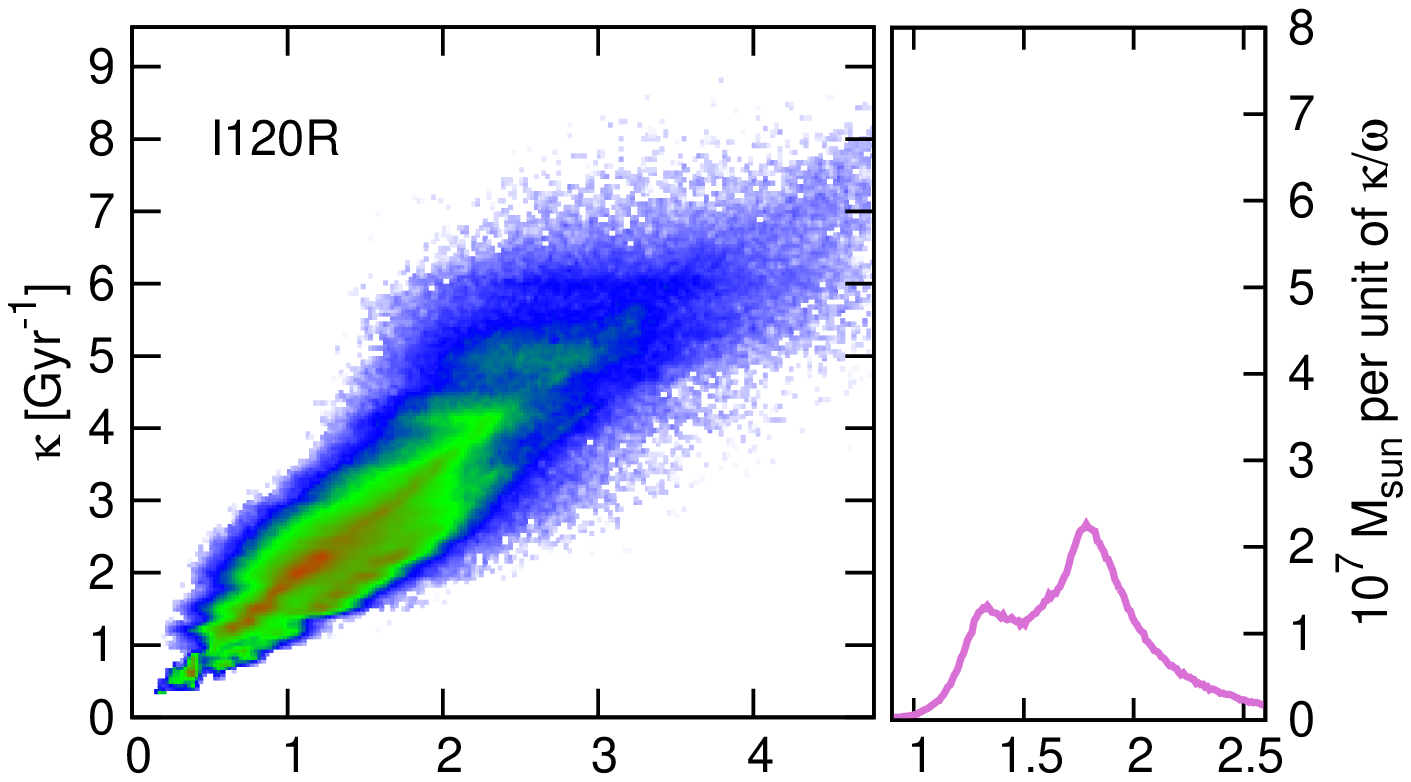}}\\
\resizebox{\hsize}{!}{\includegraphics[angle=-90]{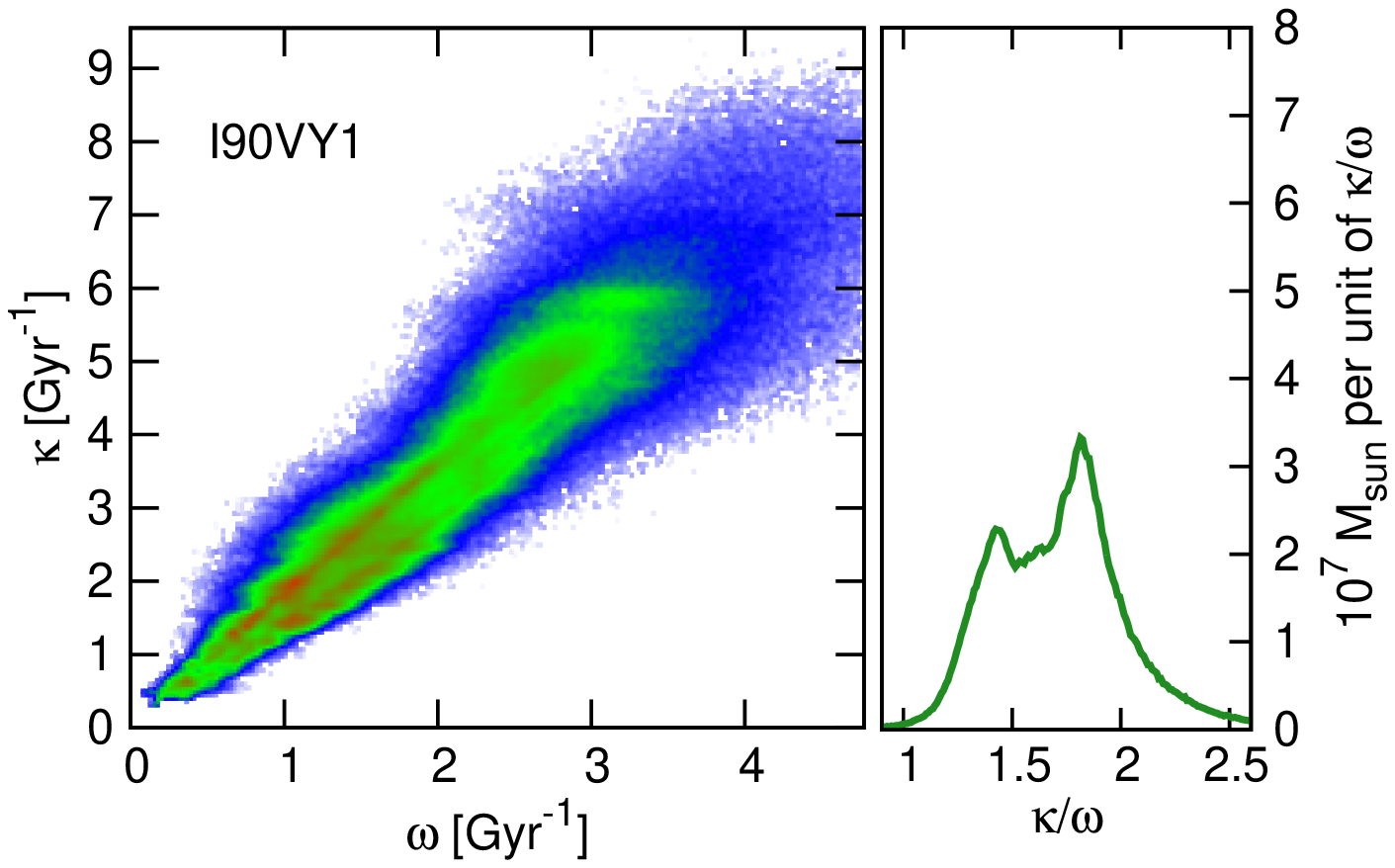}}
\caption{Left column: the distribution of the stars on long-axis tube orbits in the $\omega\kappa$
(i.e. angular vs. radial frequency) plane. Right column: the distribution of the ratio $\kappa/\omega$.
The rows from top to
bottom show results for simulations I60R, I90R, I120R and I90VY1. }
\label{fig:om}
\end{figure}
%--------------------------------------
%%%%%%%%%%%%%%%%%%%%%%%%%%%%%%%%%%%%%%%%%%%%%%%%%%%%%%%%%%%%%%%%%%%%%%%%%%%%%%%%

 \subsection{Orbital frequencies}

For the long-axis tubes we computed the average angular and radial frequency of the stellar motion in the plane
perpendicular to the major axis. The calculations were done for the period of the last 4\,Gyr of each simulation.
The angular frequency, $\omega$, was computed in each time step as $v_{\mathrm{t}}/r_{yz}$, where $v_{\mathrm{t}}$ is
the tangential part of $(v_y^2 + v_z^2)^{1/2}$ and $r_{yz}=(y^2 + z^2)^{1/2}$. We defined the radial frequency,
$\kappa$, as $1/T_{\mathrm{avr}}$, where $T_{\mathrm{avr}}$ is the average time period between two subsequent
maxima of $r_{yz}$.

The distributions of the stars in the $\omega\kappa$ plane for all four simulations are shown in the left panels
of Fig.~\ref{fig:om}. Only stellar particles with orbits classified as long-axis tubes are included.
The right panels show the distribution of the ratio $\kappa/\omega$ for the same particles.
Two branches are clearly visible in the plots. For all simulations the branches have a similar slope of
$\kappa(\omega)$: around 1.8 for the upper branch and 1.4 for the lower one. Two sub-families are known to exist
among the family of long-axis tube orbits in triaxial systems: the inner and outer long-axis tubes.
A natural expectation therefore is that the two branches correspond to these two sub-families.

In order to check this interpretation we tried to identify the inner and outer long-axis tubes using the
criteria from \citet{gaj15}. For an orbit to be classified
as an inner tube, the absolute value of the $x$-coordinate of the particle when it reaches the maximum value of
$r_{yz}$ must be larger than the corresponding value when $r_{yz}$ has a minimum, that is,
$\left|x(r_{yz,\mathrm{max}})\right| >\left|x(r_{yz,\mathrm{min}})\right| $. For the outer tubes, the inverse condition
is fulfilled: $\left|x(r_{yz,\mathrm{max}})\right| <\left|x(r_{yz,\mathrm{min}})\right| $.

\begin{figure} %[h]
\centering{}
\vspace{-2mm}
\resizebox{\hsize}{!}{\includegraphics[angle=-90]{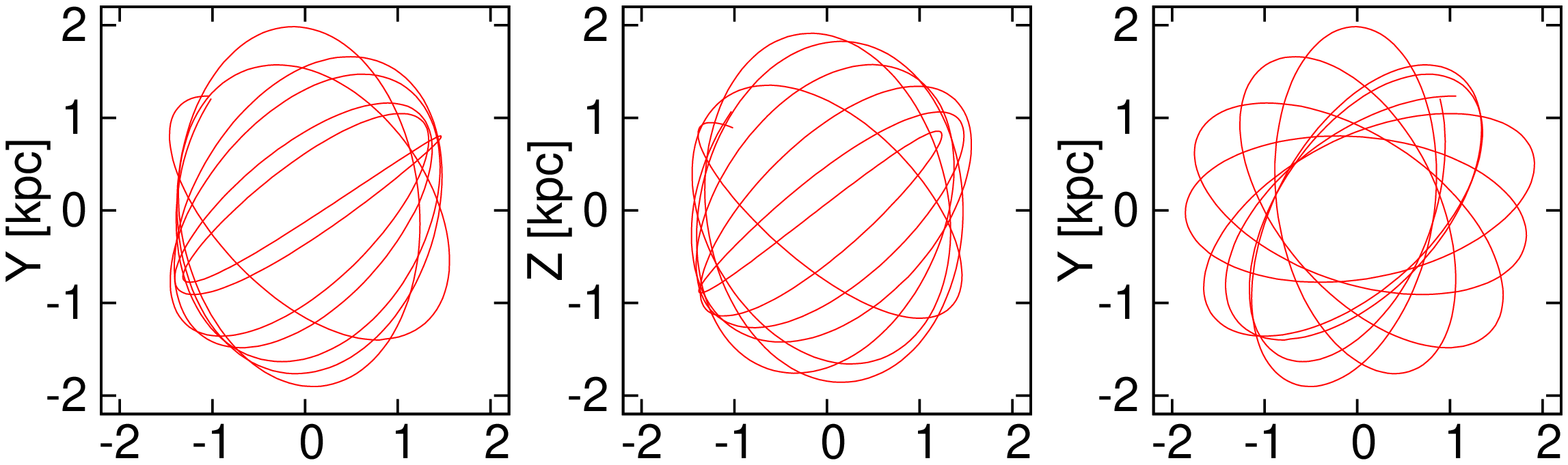}}\\
\resizebox{\hsize}{!}{\includegraphics[angle=-90]{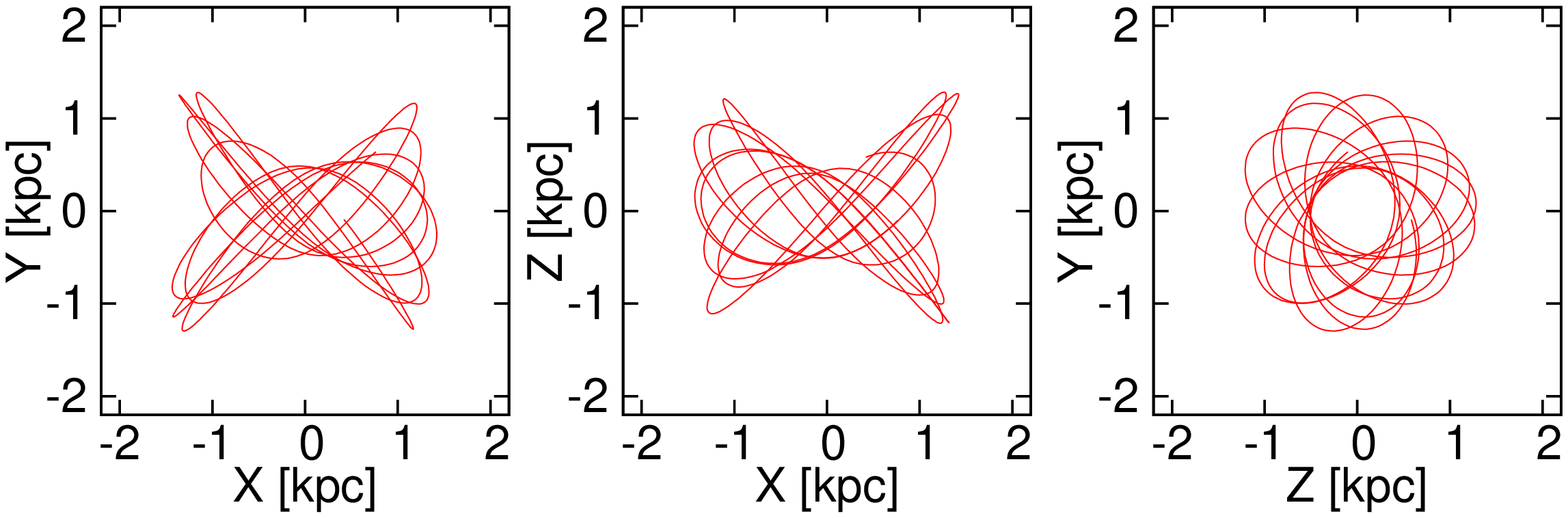}}
\caption{Examples of stellar orbits belonging to the long-axis tube family. The upper row shows three projections of the
outer long-axis tube and the lower one of the inner long-axis tube. The orbits occupy different branches of the
$\kappa(\omega)$ relation.}
\label{fig:porb}
\end{figure}

We find that indeed the outer tubes are more likely to be
found in the upper branch of the $\kappa(\omega)$ plot while the inner tubes are typically situated in the lower one.
By visual inspection of some orbits we see that this rule does not identify the tubes correctly in all cases as
particle orbits are not always regular: they are still being disturbed by perturbations related to the past merger
or by the time-dependent potential due to figure rotation.
Fig.~\ref{fig:porb} shows two
examples of orbits from the upper branch of $\kappa(\omega)$ in the top row and from the lower branch in the bottom row
that were classified as the outer and inner long-axis tubes respectively.
Clearly, the shape of the orbits is as expected for these two sub-families, i.e. the outer tubes are convex while the
inner ones are concave.

Comparing the distributions of the ratio $\kappa/\omega$ in the right panels of Fig.~\ref{fig:om} we see significant
differences among simulations. In particular, the shape of the histogram for simulation I60R is significantly different
than for the other ones. The difference is not only that there are more stars on long-axis tubes (the distribution takes larger
values) but the left peak of $\kappa/\omega \approx 1.4$ is significantly more populated than the right peak
with $\kappa/\omega \approx 1.8$. Since we have identified the left peak (lower branch) with the inner long-axis
tubes this means that the merger remnant in simulation I60R is populated more by these orbits than the outer long-axis
tubes. The situation is opposite for the remaining simulations. Since the inner tubes tend to produce more boxy
shapes this explains why the galaxy formed in I60R was much more boxy (see Fig.~\ref{fig:maps}) than the other ones.

%--------------------------------------
%--------------------------------------

\vspace{5mm}

%%%%%%%%%%%%%%%%%%%%%%%%%%%%%%%%%%%%%%%%%%%%%%%%%%%%%%%%%%%%%%%%%%%%%%%%%%%%%%%%
%%%%%%%%%%%%%%%%%%%%%%%%%%%%%%%%%%%%%%%%%%%%%%%%%%%%%%%%%%%%%%%%%%%%%%%%%%%%%%%%

\section{Mergers versus tidal stirring}

As mentioned in the Introduction, an alternative, and far better studied, scenario for the formation of dSph galaxies
in the Local Group involves their interaction with a bigger host galaxy like the Milky Way or M31. In this section
we attempt a comparison between our merger simulations and those describing such tidal stirring scenario. For this
purpose we use four simulations of dwarfs orbiting around Milky Way-sized host described in \citet{lo15}. The
initial model of the dwarfs in that paper was the same as we used for our merger progenitors and the individual
simulations differed in the angle between the angular momentum of the dwarf disk and its orbital angular momentum. The
four simulations will be referred to here by labels indicating the scenario (TS for tidal stirring) and the inclination
angle: TS-I0, TS-I90, TS-I180 and TS-I120. TS-I0 and TS-I180 correspond to the exactly prograde and exactly
retrograde orbits, respectively, while TS-I90 and TS-I270 indicate the two cases where the dwarf's disk was
perpendicular to the orbit.

\begin{figure}
\centering{}
\resizebox{\hsize}{!}{\includegraphics[angle=-90]{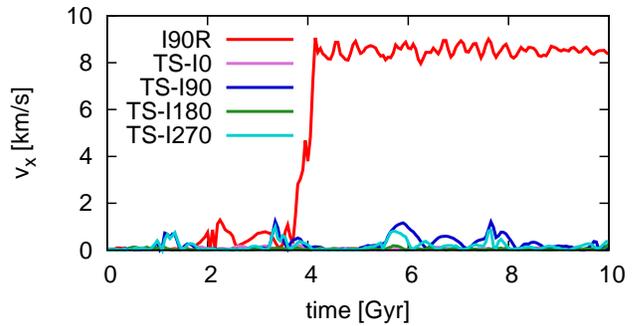}}
\caption{Evolution of the mean rotation velocity around the major axis from simulations of the tidally stirred
galaxies and Dwarf 1 of the merger simulation I90R. (Here, the colors used for other merger simulations throughout the
paper were temporarily loaned to illustrate the results of the simulations of tidal stirring.)}
\label{fig:vx}
\end{figure}

%Prolate rotation
As already discussed by \citet{lo14andii}, although using only an exactly prograde orientation of the disk, the main
and most important difference between the two scenarios seems to be the occurrence of prolate rotation which is present
in dwarfs formed by mergers and absent in the tidal stirring simulations. We are not aware of any mechanism which could
induce prolate rotation only via tidal interaction of a disky dwarf with a host galaxy. To support this statement,
in Fig.~\ref{fig:vx} we plot the evolution of the mean rotation velocity around the major axis, $v_x$, of the stellar
component of the dwarfs in the tidal stirring simulations. The measurements invariably yield very low values of this
quantity, of the order of 1 km s$^{-1}$ at most, not only for the prograde case but for the other inclinations as well.
As expected, the signal is strongest for the perpendicular orientations of the disk, i.e. in simulations TS-I90 and
TS-I270. We also note that the amount of prolate rotation is not related in any way to whether and how fast a bar
forms and evolves in these simulations.
For comparison, we also include in this Figure the same quantity measured for Dwarf\,1 of our merger simulation
I90R from Fig.~\ref{fig:rot} where the evolution of $v_x$ for all the merger simulations is shown. In the tidal
stirring simulations, the rotation around the minor axis, $v_z$, is always the dominant component of the rotation even
though it is gradually diminished and replaced by random motions as a result of the evolution \citep[see the top panel
of Figure~2 in][]{lo15}.

%--------------------------------------
%--------------------------------------

%--------------------------------------
\begin{figure}
\centering{}
\vspace{4.5mm}
\resizebox{\hsize}{!}{\includegraphics[angle=-90]{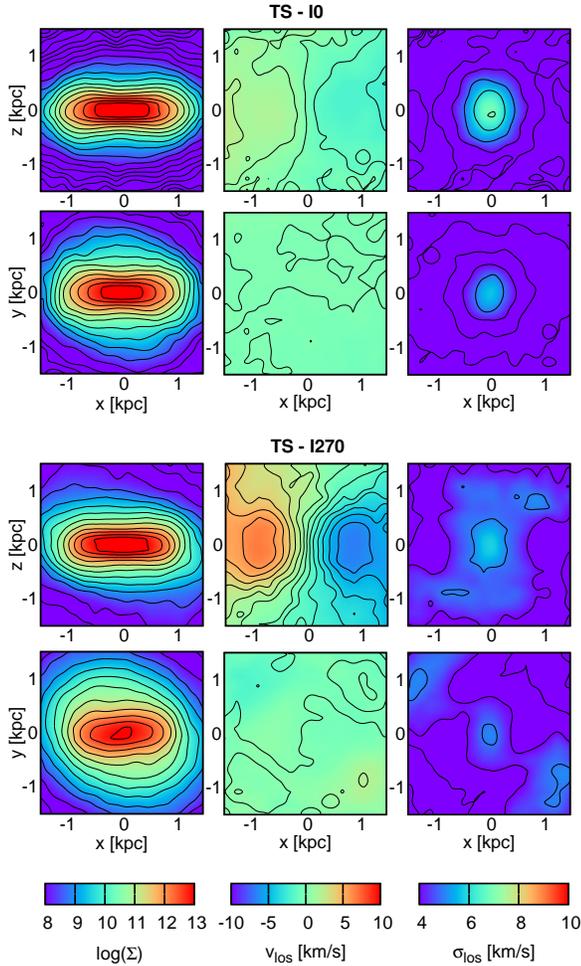}}
\vspace{-15mm}
\caption{ Maps of the surface density (left column), the mean line-of-sight velocity (middle column) and
the line-of-sight velocity dispersion (right column) for simulations TS-I0 and TS-I270 of tidally stirred galaxies.
In the first and third row the line of sight is along the intermediate axis of the galaxy, in the second and
fourth row it is along the minor axis.}
\label{fig:stmaps}
\end{figure}

%--------------------------------------

Fig.~\ref{fig:stmaps} shows the surface brightness and kinematics maps for tidally stirred dwarfs from simulations
TS-I0 and TS-270 after 6.5\,Gyr of evolution (i.e. at the fourth apocenter) viewed along the intermediate and minor axis
($y$ and $z$, respectively). The maps are analogous to those shown in Fig.~\ref{fig:maps} for the merger remnants, we
kept the same range of the color scale as well as distances between the isocontours for the corresponding
quantities. Only the displayed area is smaller here because the tidally stirred dwarfs are
diminished in size at this late stage of evolution due to tidal stripping. The maps of the mean velocity (middle
columns) clearly show that the rotation, if present, is seen always along the major axis of the projected galaxy so the
stars rotate around the minor axis.
%velocity dispersion
The velocity dispersion of the tidally stirred dwarfs (right columns of Fig.~\ref{fig:stmaps}) is generally lower and
more regular than in our merger remnants where the dispersion maintains a more complicated structure even many Gyr after
the merger. In order to detect these irregularities one would need a good spatial resolution for the kinematics and we
comment on this issue further in the next section.

%It is important to note that the velocity dispersion of stirred galaxies is not completely smooth for all the cases at
%all times and thus this characteristic must be treated with caution.

%Boxiness
The surface density maps for the tidally stirred dwarfs in the left columns of Fig.~\ref{fig:stmaps} show the remnant
bar in the center. In comparison with these, the merger remnants have overall more regular shapes with more boxy
isocontours. However, the boxiness is suppressed for nonradial mergers and larger initial angles between angular
momenta of the progenitor discs. In addition, the bar induced during the tidal stirring can make the isophotes look
boxy as well if only the brightest central part of the galaxy was available for observation, so this feature probably
cannot be used as an indicator of the merger origin. In the case of low surface brightness galaxies, such as
And\,II, the boxy shape would be particularly challenging to detect.

%Density profile
In principle, one more feature that could discriminate between the tidally stirred and merged objects, as advertised by
\citet{tom15}, is the density profile in the outskirts. It has been shown \citep[e.g.][]{lo13, lo15} that tidally
stirred galaxies develop a flattening of the density profile due to the transition between the stars which are bound to
the dwarf and the tidal tails. However, the presence and detectability of this transition depends on the number of
issues. First, the transition in the form of the break radius in the stellar density profile is not seen in
dwarfs in non-prograde orbital configurations \citep[see Figure 6 in][]{lo15}. Second, the tails will not be visible if
they are aligned with the line of sight of the observer and this unfortunately is the case for most of the
time for dwarfs on different orbits \citep{kli09, lo13}. Third, the tidal tails are usually very faint and
have been convincingly detected only for the Sgr dwarf. Moreover, tails produced by the merger can, at some stages,
mimic this transition. Furthermore, one cannot rule out the case of the merger remnants being subsequently tidally
stirred by the host galaxy.

\section{Towards a complete model of And II evolution}

In this section we compare the results of our merger simulations to the real data for And II and discuss the possible
scenario for its formation. In order to reproduce the observational data, we require the simulated merger remnant
to show: (1) the rotation along the minor axis at the level of 10\,km\,s$^{-1}$ at the maximum, which should be
comparable to the maximum value of the line-of-sight velocity dispersion, (2) no significant rotation along the major
axis, (3) the same kinematics for stars originating from both progenitors, (4) the shape of rather low ellipticity. The
observational data cover the area of about 2 kpc in radius.

\begin{figure}
\centering{}
\resizebox{\hsize}{!}{\includegraphics[angle=-90]{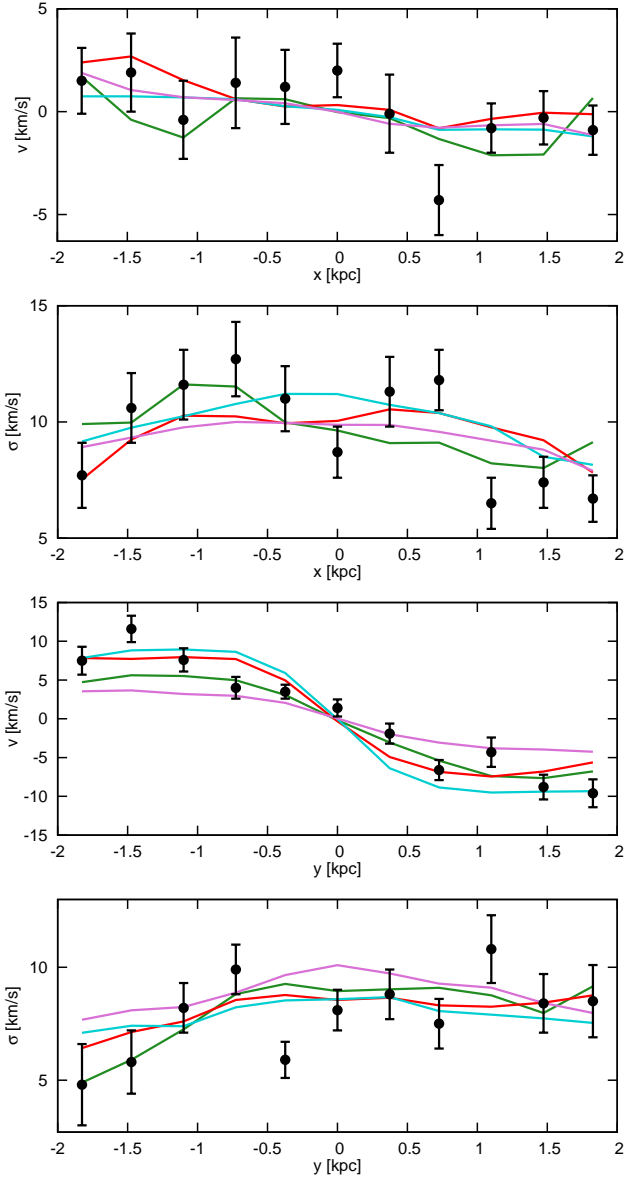}}
\caption{Comparison between the observed \citep[black points;][]{ho12} and simulated (color lines) kinematics of
And\,II. The panels show, from top to bottom, the mean line-of-sight velocity and line-of-sight velocity dispersion
along the major axis of the observed galaxy ($x$) and the mean line-of-sight velocity and line-of-sight dispersion
along the minor projected axis ($y$). For simulated galaxies, the line of sight is inclined with respect to the
minor axis of the stellar ellipsoid by [27,167] deg for simulation I90VY1 (4.6\,Gyr; green lines), [18,166]
for I60R (5.65\,Gyr; cyan), [9,32] for I90R (4.7\,Gyr; red), and [32,158] for I120R (7.1\,Gyr; magenta).}
\label{fig:bestkin}
\end{figure}

The measured kinematics of And\,II \citep{ho12} shows significant variability, while the
equivalent values taken from the final outputs of the simulations are rather smooth if they are derived using all
stellar particles along the major/minor projected axis about 4\,kpc long. In this case even the fluctuations
of the dispersion are averaged out for the final outputs. Therefore, in order to reproduce the observed kinematics we
explored the kinematics from all simulation outputs (after the merger) as well as different lines of sight along
which these outputs are `observed'. Best matches of the observed kinematics (in terms of the lowest $\chi^2$) for all
simulations are shown in Fig.~\ref{fig:bestkin}. In this comparison we again exclude the non-radial simulation I90VY2
which failed to produce significant amount of prolate rotation. The sizes of the bins for the simulated data correspond
to those used in observations. The best matches come from times 4.6--7.1\,Gyr since the beginning of the simulations.
In all cases
we also verified that for the chosen line of sight the ellipticity of the galaxy is in the range of about 0.1--0.2, as
required by the data \citep{mci06, ho12, sa15}.

%%%%%%%%%%%%%%%%%%%%%%%%%%%%%%%%%%%%%%%%%%%%%%%%%%%%%%%%%%%%%%%%%%%%%%%%%%%%%%%%
%--------------------------------------

%--------------------------------------

%--------------------------------------
\begin{figure}
\centering{}
\resizebox{\hsize}{!}{\includegraphics[angle=-90]{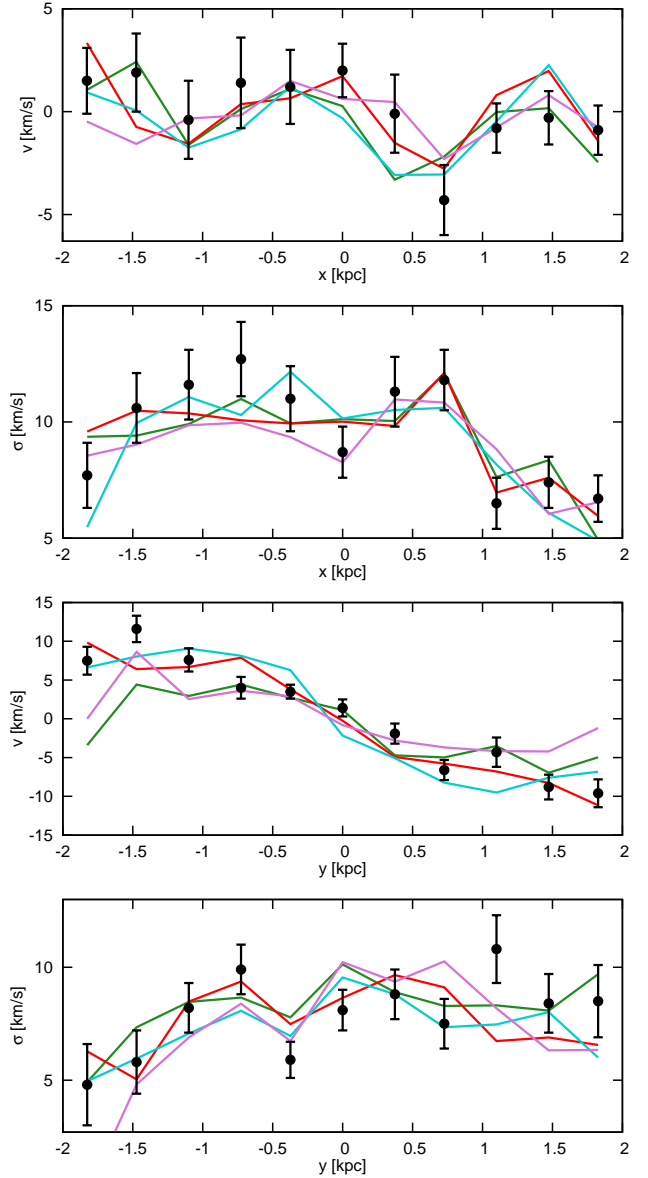}}
\caption{The same as Fig.~\ref{fig:bestkin} but for a subsample of about 500 stellar particles from the final
outputs (at 10\,Gyr) of the simulations. The line of sight is inclined with respect to the minor axis of the
stellar ellipsoid by [45,127] deg for I90VY1, [18,114] for I60R, [27,126] for I90R, and [36,137] for I120R.}
\label{fig:kin531}
\end{figure}
%--------------------------------------

%--------------------------------------
\begin{figure*}
\centering{}
\resizebox{.95\hsize}{!}{\includegraphics[angle=-90]{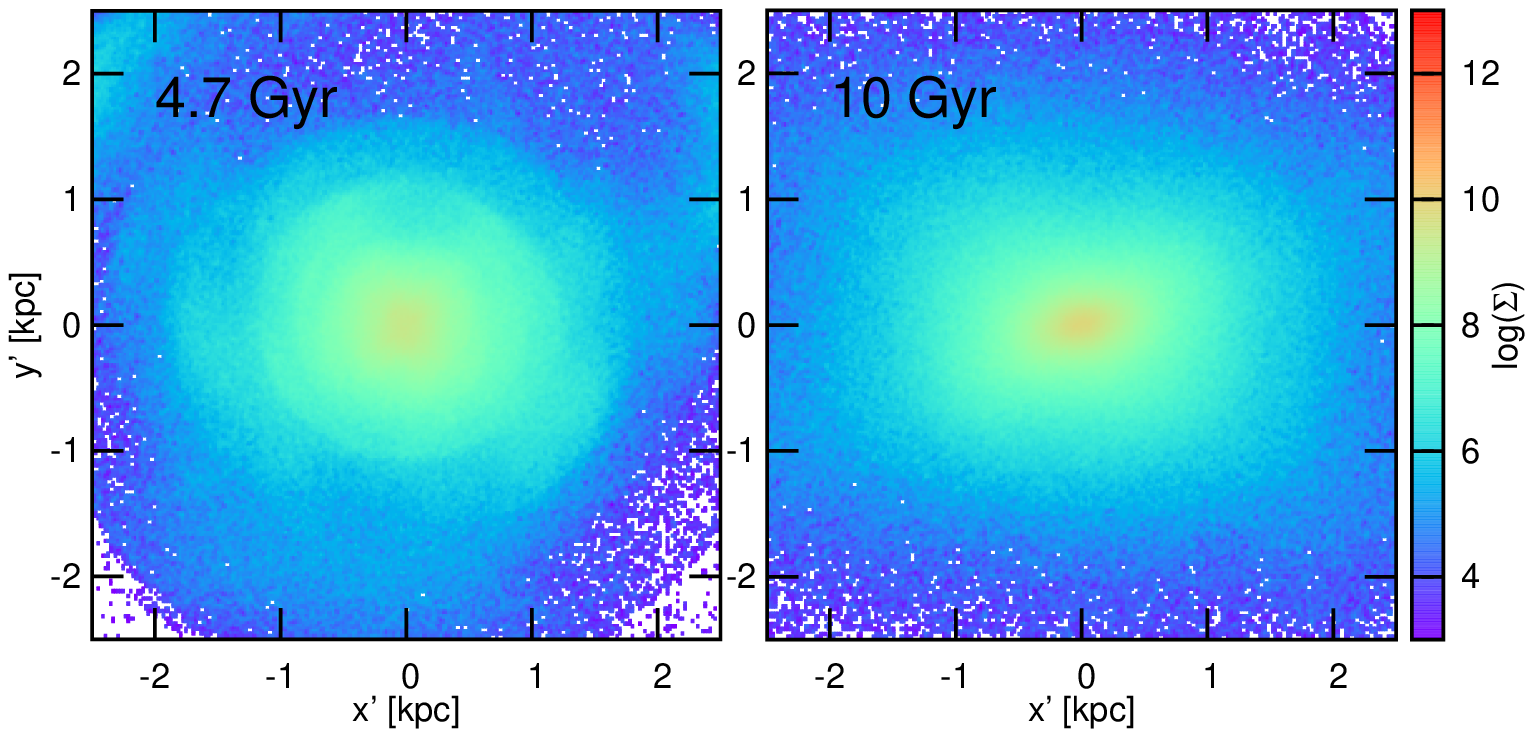}}
\caption{Surface density of the stellar component in simulation I90R at 4.7\,Gyr (left panel) and 10\,Gyr (right
panel). The line of sight is inclined with respect to the minor axis of the stellar ellipsoid by [9,32] deg
for the output at 4.7\,Gyr and by [27,126] deg for the output at 10\,Gyr.}
\label{fig:sb094-200}
\end{figure*}
%--------------------------------------

In general, earlier stages of the merger can reproduce the fluctuations in the measured kinematics better. Shortly after
the merger, the galaxy is unrelaxed with many shells and streams visible that formed as a result of the collision. An
example of the merger remnant image corresponding to the output at 4.7 Gyr with the best-fitting kinematics for
simulation I90R is shown in the left panel of Fig.~\ref{fig:sb094-200}.

However, we are able to reproduce the fluctuations in the kinematics even for the final outputs corresponding to times
a few Gyr after the merger if we recall that the real data are based on a much lower number of stars. For this
comparison, we pick a subsample of about 500 stellar particles which match the number of stars used to derive the
kinematics of And\,II in \citet{ho12}. Again we use the same binning and select the stellar particles so that their
distribution roughly covers the area used in the observations. In all other aspects, the choice of the
particles is random. We repeat the subsample selection several hundred times for different lines of sight and find
the ones that match the data best.

Examples of good matches of the simulated data to the observed kinematics are
shown in Fig.~\ref{fig:kin531} for cases with ellipticity around 0.1--0.2. This proves that the observed fluctuations
could be still consistent with the relaxed stage of the merger remnant. An image of such a relaxed merger remnant in
a late stage is shown in the right panel of Fig.~\ref{fig:sb094-200} for simulation I90R at 10 Gyr where it can
be directly compared to the analogous image at 4.7 Gyr in the left panel. These outputs and lines of sight correspond
to the kinematics shown with red lines in Fig.~\ref{fig:bestkin} (for 4.7\,Gyr) and Fig.~\ref{fig:kin531} (for
10\,Gyr).

%description of the possible scenario

Besides the measured shape and kinematics, additional constraints on the possible scenario for the formation of And II
come from its recently measured star formation history \citep{var14,wei14,ga15}. The measurements point to the
existence of the intermediate-age stellar population dated 5--8 Gyr ago. Although we do not model star formation processes
here, given the comparisons discussed above and these additional constraints we may speculate on the possible general
scenario for the formation of And II.

And II most likely formed as a result of a merger between two disky gas-rich dwarf galaxies of similar mass and disk
properties. The disks must be inclined by $90\pm30$ deg in order to possess significant components of angular
momentum along the merger axis. The two galaxies approached each other on a nearly radial orbit with a rather low
relative velocity forming a bound pair. The existence of such pairs of subhalos has been confirmed in recent studies
of simulations reproducing environments similar to the Local Group \citep{kli10, kaz11, dea14, wet15}. Since the
occurrence of such groups is more likely outside the virial radius of big galaxies, the merger probably occurred
before the galaxies were accreted by M31, a few Gyr ago. The merger induced the secondary burst of star formation,
mainly in the center of the merger remnant, converting some of the gas into stars in a similar fashion as in the
well-studied mergers between normal-size galaxies \citep[e.g.][]{cox06, hof10}. The new stars gave rise to the distinct
stellar population with a more concentrated density profile that can be identified with the
intermediate-age stellar population seen in the data. These younger stars should contribute about 30\%
of the stellar mass in the central region \citep{ga15}.

Further evolution of the merger remnant remains an open question. However, since we do not see any young stars in And
II and no gas is detected in it at present, we may suppose that the star formation ceased a few Gyr ago,
either quenched by internal processes such as supernova feedback or by environmental effects such as ram pressure
stripping, if the merger remnant was accreted by M31. The present distance of And II from M31 is about 180 kpc and
their relative line-of-sight velocity is of the order of 100 km s$^{-1}$ i.e. low enough for And II to be a
satellite of M31, but no strong constraints are available on its orbit. However, \citet{st13} find a solution for the
Local Group galaxy orbits (see their Figure 13) where And II has experienced a close passage near M31 which may have
stripped its gas. Nevertheless, this interaction could not be too strong so that And II could preserve its
prolate rotation and probably did not lead to any new starburst.
Cases of accretion of merger remnants by bigger hosts have been also confirmed in the simulations of
the Local Group and Milky Way environments \citep{kaz11, tom15}.

The scenario we sketched here can be developed into a full model of a gas-rich merger
only by incorporating gas dynamics and star
formation. The outcome of such simulations will depend on a number of parameters like the threshold
for star formation and the amount of feedback. These parameters can in principle be adjusted to reproduce the observed
mass fraction of the newer stellar population. We expect that such a model will be able to preserve
the prolate rotation of the old stellar population but it remains to be seen what will be the kinematics of
the younger stars. A more complete scenario for the formation of And II will be described in detail by
S. Fouquet et al. (in preparation).

%%%%%%%%%%%%%%%%%%%%%%%%%%%%%%%%%%%%%%%%%%%%%%%%%%%%%%%%%%%%%%%%%%%%%%%%%%%%%%%%
%%%%%%%%%%%%%%%%%%%%%%%%%%%%%%%%%%%%%%%%%%%%%%%%%%%%%%%%%%%%%%%%%%%%%%%%%%%%%%%%
\section{Summary and conclusions}

In this work we studied the origin of prolate rotation in dSph galaxies forming via mergers of two disky dwarfs.
Our work was motivated by the discovery of such type of rotation in the dSph galaxy And\,II.
\citet{lo14andii} demonstrated that
this kind of rotation can indeed result from a merger between two disky dwarf galaxies. They proposed a detailed,
but rather fine-tuned scenario with very special initial conditions. Here we showed that these initial conditions
can be varied to some extent and can still reproduce the main observational properties of And\,II.

Masses of the luminous parts of the dwarfs were chosen so that their combined luminosity is similar
to what is measured for And\,II. The dark matter components were adjusted so that the kinematics of the remnants
matches the observations. In addition, \citet{lo14andii} showed that the different density profiles of the
two stellar populations could be reproduced by adopting
different initial scale-lengths of the disks of the dwarfs. We did not address this issue in the present paper and for
simplicity assumed that the two progenitors are identical in terms of structural parameters and masses.

The evolution of the three radial mergers we considered is straightforward to interpret.
In these simulations the initial conditions are symmetrical for
both dwarfs. The initial angle between the angular momenta of the dwarf disks determines how much of the disk rotation
is transformed into the prolate rotation of the merger product. With the lower value of the angle, corresponding to
the more vertical orientation of the disks, we get more prolate rotation as well as a more boxy shape of
the surface density of the galaxy.
The dispersion and, more importantly, the rotation in simulation I120R is not high enough to match the
data while I90R and I60R both reproduce the line-of-sight kinematics of And\,II satisfactorily.

From non-radial simulations (I90VY1 and I90VY2) we see that the perpendicular component of the relative velocity,
corresponding to the orbital angular momentum, cannot
be increased too much. Both simulations have the radial component of the relative velocity of 8\,km\,s$^{-1}$.
In simulation I90VY1, with the
perpendicular velocity component of 1\,km\,s$^{-1}$, the merger remnant still
roughly meets the observation constraints (although with less rotation and smaller
dispersion than observed). The remnant of simulation I90VY2, with the perpendicular velocity component of
2\,km\,s$^{-1}$, does not reproduce the data well:
it has very small rotation and the kinematics of stars originating from different dwarfs is significantly different.
We note that for non-radial mergers the initial conditions were set up so that the merger is prograde for
one progenitor and retrograde for the other one. In combination with a larger orbital angular momentum, this
resulted in a very different evolution in simulation I90VY2 compared to I90VY1.

All remnants showing significant prolate rotation have similar orbital structure, dominated by box orbits in the center
of the galaxy and by the long-axis tubes in the outer parts. However, there are subtle differences depending on the
initial conditions: the transition from the box orbits to the long-axis tubes takes place at smaller radii for
remnants with stronger prolate rotation. In addition, the remnant with the highest rotation and
the most boxy shape (I60R) turns out to possess a noticeably bigger contribution of inner versus outer long-axis
tubes than any other remnant.

We conclude that prolate rotation may be produced in mergers with a variety of initial conditions.
The inclination of the disks with respect to the orbit must be significant in order to provide enough angular momentum
in the remnants but can be varied by a few tens of degrees, producing remnants of slightly different kinematics and
shape. Some orbital angular momentum during the merger, up to a factor of a few
of the intrinsic angular momentum of the merging disks, is also allowed and does not destroy the symmetry between
the two components of the remnant.

We compared the predicted properties of our merger remnants to those characteristic of galaxies evolving in the
vicinity of bigger host and affected by tidal stirring. Although a number of features, such as the boxiness of the
shape, irregularities in the kinematics and strong tidal extensions, could possibly help us to distinguish the dSph
galaxies formed in the two scenarios, they are generally not accessible with present observational techniques. We
propose that the only reliable feature that can act as a fingerprint of the past merger is the presence of prolate
rotation such as the one observed in And II. Such type of rotation can not be induced via tidal stirring at a
significant level even for a variety of orbital configurations.

We have also compared our merger remnants to the real data available for And II. We find that in order to reproduce the
data the merger could not have happened recently, but rather a few Gyr ago. This statement is supported by the fact
that the surface brightness distribution in And II is rather regular without obvious distortions and the need to be
consistent with the recently measured star formation history of the galaxy which points to the presence of an
intermediate-age, rather than a young, stellar population. The subsequent evolution of And II should have been rather
quiescent in order to preserve prolate rotation, although it may have interacted with M31, which led to the stripping
of its gas.

%%%%%%%%%%%%%%%%%%%%%%%%%%%%%%%%%%%%%%%%%%%%%%%%%%%%%%%%%%%%%%%%%%%%%%%%%%%%%%%%
%%%%%%%%%%%%%%%%%%%%%%%%%%%%%%%%%%%%%%%%%%%%%%%%%%%%%%%%%%%%%%%%%%%%%%%%%%%%%%%%

\acknowledgments
This research was supported in part by PL-Grid Infrastructure, by the Polish National Science Centre under grant
2013/10/A/ST9/00023 and by the project RVO:67985815. IE is grateful for the hospitality of the Copernicus Center in
Warsaw during her visits. Access to computational resources provided by the MetaCentrum under the program LM2010005 and
the CERIT-SC under the program Centre CERIT Scientific Cloud, part of the Operational Program Research and Development
for Innovations, Reg. no. CZ.1.05/3.2.00/08.0144, is greatly appreciated. We thank L. Widrow for providing procedures to
generate $N$-body models of galaxies for initial conditions. We are grateful to the anonymous referee for comments
that helped us to significantly improve the paper.

%%%%%%%%%%%%%%%%%%%%%%%%%%%%%%%%%%%%%%%%%%%%%%%%%%%%%%%%%%%%%%%%%%%%%%%%%%%%%%%%
\bibliographystyle{apj}

\begin{thebibliography}{}
\expandafter\ifx\csname natexlab\endcsname\relax\def\natexlab#1{#1}\fi

\bibitem[{{Amorisco} {et~al.}(2014){Amorisco}, {Evans}, \& {van de Ven}}]{am14}
{Amorisco}, N.~C., {Evans}, N.~W., \& {van de Ven}, G. 2014, \nat, 507, 335

\bibitem[{{Coleman} {et~al.}(2005){Coleman}, {Da Costa}, {Bland-Hawthorn} \& {Freeman}}]{col05}
{Coleman}, M. G., {Da Costa}, G. S., {Bland-Hawthorn}, J., \& Freeman, K. C. 2005, \aj, 129, 1443

\bibitem[{Cox et al.}(2006)]{cox06} Cox, T. J., Dutta, S. N., Di Matteo, T., et al. 2006, ApJ, 650, 791

\bibitem[{{Crnojevi{\'c}} {et~al.}(2014){Crnojevi{\'c}}, {Sand}, {Caldwell},
  {Guhathakurta}, {McLeod}, {Seth}, {Simon}, {Strader}, \& {Toloba}}]{crn14}
{Crnojevi{\'c}}, D., {Sand}, D.~J., {Caldwell}, N., {et~al.} 2014, \apjl, 795,
  L35

\bibitem[{{de Zeeuw}(1985)}]{dez85}
{de Zeeuw}, T. 1985, \mnras, 216, 273

\bibitem[{{Deason} {et~al.}(2014){Deason}, {Wetzel}, \&
  {Garrison-Kimmel}}]{dea14}
{Deason}, A., {Wetzel}, A., \& {Garrison-Kimmel}, S. 2014, \apj, 794, 115

\bibitem[{{Gajda} {et~al.}(2015){Gajda}, {{\L}okas}, \& {Wojtak}}]{gaj15}
{Gajda}, G., {{\L}okas}, E.~L., \& {Wojtak}, R. 2015, \mnras, 447, 97

\bibitem[{Gallart et al.}(2015)]{ga15} Gallart, C.; Monelli, M.; Mayer, L., et al. 2015, ApJ, 811, L18

\bibitem[{{Geha} {et~al.}(2005){Geha}, {Guhathakurta}, \& {van der
  Marel}}]{geh05}
{Geha}, M., {Guhathakurta}, P., \& {van der Marel}, R.~P. 2005, \aj, 129, 2617

\bibitem[{{Ho} {et~al.}(2012){Ho}, {Geha}, {Munoz}, {Guhathakurta}, {Kalirai},
  {Gilbert}, {Tollerud}, {Bullock}, {Beaton}, \& {Majewski}}]{ho12}
{Ho}, N., {Geha}, M., {Munoz}, R.~R., {et~al.} 2012, \apj, 758, 124

\bibitem[{{Ho} {et~al.}(2015){Ho}, {Geha}, {Tollerud}, {Zinn}, {Guhathakurta},
  \& {Vargas}}]{ho15}
{Ho}, N., {Geha}, M., {Tollerud}, E., {et~al.} 2015, \apj, 798, 77

\bibitem[{Hoffman et al.}(2010)]{hof10} Hoffman, L., Cox, T. J., Dutta, S., \& Hernquist, L. 2010, ApJ, 723, 818

\bibitem[{Kazantzidis et al.}(2011)]{kaz11} Kazantzidis, S., {\L}okas, E. L., Mayer, L., Knebe, A.,
	\& Klimentowski, J. 2011, \apj, 740, L24

\bibitem[{Klimentowski et al.}(2009)]{kli09} Klimentowski, J., {\L}okas, E. L., Kazantzidis, S.,
	et al. 2009, MNRAS, 400, 2162

\bibitem[{{Klimentowski} {et~al.}(2010){{Klimentowski}, {\L}okas}, {Knebe},
  {Gottl\"ober}, {Martinez-Vaquero}, {Yepes}, \& {Hoffman}}]{kli10}
{Klimentowski}, J., {{\L}okas}, E.~L., {Knebe}, A., {et~al.}
  2010, \mnras, 402, 1899

\bibitem[{{{\L}okas} {et~al.}(2013){{\L}okas}, {Gajda}, \&
  {Kazantzidis}}]{lo13} {\L}okas, E. L., Gajda, G., \& Kazantzidis, S. 2013, MNRAS, 433, 878

\bibitem[{{{\L}okas} {et~al.}(2014{\natexlab{a}}){{\L}okas}, {Ebrova}, {del Pino}, \&
  {Semczuk}}]{lo14andii}
{{\L}okas}, E.~L., {Ebrova}, I., {del Pino}, A., \& {Semczuk}, M.
  2014{\natexlab{a}}, \mnras, 445, L6

\bibitem[{{{\L}okas} {et~al.}(2014{\natexlab{b}}){{\L}okas}, {Athanassoula},
  {Debattista}, {Valluri}, {del Pino}, {Semczuk}, {Gajda}, \&
  {Kowalczyk}}]{lo14bar}
{{\L}okas}, E.~L., {Athanassoula}, E., {Debattista}, V.~P., {et~al.}
  2014{\natexlab{b}}, \mnras, 445, 1339

\bibitem[{{{\L}okas} {et~al.}(2015){{\L}okas}, {Semczuk}, {Gajda}, \&
  {D'Onghia}}]{lo15}
{{\L}okas}, E.~L., {Semczuk}, M., {Gajda}, G., \& {D'Onghia}, E.
  2015, \apj, 810, 100

\bibitem[{{Mart{\'{\i}}nez-Delgado} {et~al.}(2012){Mart{\'{\i}}nez-Delgado},
  {Romanowsky}, {Gabany}, {Annibali}, {Arnold}, {Fliri}, {Zibetti}, {van der
  Marel}, {Rix}, {Chonis}, {Carballo-Bello}, {Aloisi}, {Macci{\`o}},
  {Gallego-Laborda}, {Brodie}, \& {Merrifield}}]{md12}
{Mart{\'{\i}}nez-Delgado}, D., {Romanowsky}, A.~J., {Gabany}, R.~J., {et~al.}
  2012, \apjl, 748, L24

\bibitem[{McConnachie \& Irwin}(2006)]{mci06} McConnachie, A. W., \& Irwin, M. 2006, MNRAS, 365, 1263

\bibitem[{{McConnachie} {et~al.}(2007){McConnachie}, {Arimoto}, \&
  {Irwin}}]{mcc07}
{McConnachie}, A.~W., {Arimoto}, N., \& {Irwin}, M. 2007, \mnras, 379, 379

\bibitem[{{Navarro} {et~al.}(1997){Navarro}, {Frenk}, \& {White}}]{nfw97}
{Navarro}, J.~F., {Frenk}, C.~S., \& {White}, S.~D.~M. 1997, \apj, 490, 493

\bibitem[{{Paudel} {et~al.}(2015){Paudel}, {Duc}, \& {Ree}}]{pau15}
{Paudel}, S., {Duc}, P.~A., \& {Ree}, C.~H. 2015, \aj, 149, 114

\bibitem[{Salomon et al.}(2015)]{sa15} Salomon, J.-B., Ibata, R. A., Martin, N. F., \& Famaey, B.
	2015, MNRAS, 450, 1409

\bibitem[{{Schwarzschild}(1982)}]{schw82}
{Schwarzschild}, M. 1982, \apj, 263, 599

\bibitem[{{Schwarzschild}(1993)}]{schw93}
---. 1993, \apj, 409, 563

\bibitem[{Shaya \& Tully}(2013)]{st13} Shaya, E. J., \& Tully, R. B. 2013, MNRAS, 436, 2096

\bibitem[{{Springel}(2005)}]{sp05}
{Springel}, V. 2005, \mnras, 364, 1105

\bibitem[{{Tomozeiu} {et~al.}(2015){Tomozeiu}, {Mayer}, \& {Quinn}}]{tom15}
{Tomozeiu}, M., {Mayer}, L., \& {Quinn}, T. 2015, arXiv:1506.02140

\bibitem[{{Vargas} {et~al.}(2014){Vargas}, {Geha}, \& {Tollerud}}]{var14}
{Vargas}, L.~C., {Geha}, M.~C., \& {Tollerud}, E.~J. 2014, \apj, 790, 73

\bibitem[{{Weisz} {et~al.}(2014){Weisz}, {Dolphin}, {Skillman}, {Holtzman},
  {Gilbert}, {Dalcanton}, \& {Williams}}]{wei14}
{Weisz}, D.~R., {Dolphin}, A.~E., {Skillman}, E.~D., {et~al.} 2014, \apj, 789, 147

\bibitem[{Wetzel et al.}(2015)]{wet15} Wetzel, A. R., Deason, A. J., \& Garrison-Kimmel, S. 2015, ApJ, 807, 49

\bibitem[{{Widrow} \& {Dubinski}(2005)}]{wid05}
{Widrow}, L.~M., \& {Dubinski}, J. 2005, \apj, 631, 838

\bibitem[{{Widrow} {et~al.}(2008){Widrow}, {Pym}, \& {Dubinski}}]{wid08}
{Widrow}, L.~M., {Pym}, B., \& {Dubinski}, J. 2008, \apj, 679, 1239

\end{thebibliography}

\end{document}